\documentclass[11pt]{article}

\usepackage{hyperref,amsfonts,amsmath,amssymb,bm,a4wide,cite,subfigure,graphicx}

\usepackage{mathabx}

\numberwithin{equation}{section}

\begin{document}

\title{\textbf{Quantum field theory treatment of oscillations of Dirac neutrinos
in external fields}}

\author{Maxim Dvornikov\thanks{maxim.dvornikov@gmail.com}
\\
\small{\ Pushkov Institute of Terrestrial Magnetism, Ionosphere} \\
\small{and Radiowave Propagation (IZMIRAN),} \\
\small{108840 Moscow, Troitsk, Russia}}

\date{}

\maketitle

\begin{abstract}
We study neutrino oscillations in external fields using the approach
based on the quantum field theory (QFT). Neutrinos are virtual particles
in this formalism. Neutrino mass eigenstates are supposed to be Dirac fermions. We consider two cases of external fields: the neutrino
electroweak interaction with background matter and the interaction
with an external magnetic field owing to the presence of the transition
magnetic moment. The formalism used involves the dressed propagators
of mass eigenstates in external fields. In the matter case, finding
of these propagators for Dirac neutrinos has certain difficulties
compared to the Majorana particles considered previously. These difficulties
are overcome by regularizing the effective potential of the neutrino
interaction with matter. The QFT formalism application to the spin-flavor
precession also encounters certain peculiarities in the Dirac case
compared to the Majorana one. They are related to the observability
of right polarized Dirac neutrinos. We derive the matrix elements
and the probabilities for Dirac neutrinos interacting with both types
of external fields. In case of the spin-flavor precession, we obtain
the small QFT contribution to the probabilities in addition to the
prediction of the quantum mechanical approach.
\end{abstract}

\section{Introduction}\label{sec:INTR}

Studies of neutrinos open a unique possibility to explore physics
beyond the standard model. Indeed, these particles were established
in numerous experiments (see, e.g., Ref.~\cite{Abe25}) to possess
nonzero masses and mixing between different neutrino types. These
neutrino properties lead to conversions of neutrino flavors, which
are called neutrino flavor oscillations. Flavor oscillations can happen
while a neutrino beam propagates even in vacuum.

Interactions with various external fields were established theoretically
to affect neutrino oscillations. For example, the electroweak interaction
of neutrinos with other background fermions can result in the significant
amplification of the transition probability of neutrino oscillations,
named the Mikheyev--Smirnov--Wolfenstein effect~\cite{Wol78,MikSmi85}.
It is believed to be the most plausible explanation of the solar neutrino
problem~\cite{CheXu25}.

Besides the electroweak interactions, neutrinos can interact with
external electromagnetic fields owing to the presence of nonzero magnetic
moments~\cite{LeeSch77} which are not excluded experimentally. Note
that neutrino magnetic moments have purely anomalous origin~\cite{DvoStu04}
in contrast to charged leptons. Both a single and different neutrino
flavors can be involucrated in an electromagnetic interaction since,
besides diagonal magnetic moments, transition ones are also possible.
If a beam of neutrinos, having arbitrary types of magnetic moments,
propagates trough a magnetic field, both a helicity flip and the flavor
change can happen. This process is called the neutrino spin-flavor
precession~\cite{LimMar88,Akh88}.

Nowadays, all fundamental fermions, discovered experimentally, are
Dirac particles. That is, an antiparticle is different from a particle.
Neutrinos are hoped to be Majorana fermions for which particles coincide
with antiparticles. At least, the belief to the Majorana nature of
neutrinos results in the theoretical explanation of the smallness
of active neutrino masses~\cite{Kin04}. We should stress that the
issue whether neutrinos are Dirac or Majorana particles is still open
in spite of significant experimental efforts (see, e.g., Ref.~\cite{Agr25}).

The electromagnetic properties are different for Dirac and Majorana
neutrinos. Dirac neutrinos can have arbitrary magnetic moments, whereas
only transition magnetic moments are allowed for Majorana neutrinos.
The electrodynamics of neutrinos is reviewed recently in Ref.~\cite{Giu25}.

Historically, neutrino oscillations were described within the quantum
mechanical (QM) approach. Despite the QM treatment provides a satisfactory
description of neutrino oscillations in almost all situations, this
kind of formalism reveals certain shortcomings which are listed, e.g.,
in Ref.~\cite{NauNau20}. Therefore, an approach for neutrino oscillations,
free of self-contradictions, should be based on the quantum field
theory (QFT). The QFT formalism, where neutrinos are virtual particles,
was proposed in Refs.~\cite{Kob82,GiuKimLee93,GriSto96}.

The key issue in this approach is using the propagators of neutrino
mass eigenstates in the calculation of the matrix elements. If neutrinos
propagate between a source and a detector in vacuum, the propagator
has a well-known form. It allows one to calculate the transition
probability for neutrino oscillations in vacuum. The generalization
of this formalism for the description of neutrino oscillations in
external fields is nontrivial since, as we shall see shortly, neutrino
mass eigenstates are mixed in these situations. One can apply the
QFT formalism in question straightforwardly only in a particular case
when an external field is diagonal in the mass eigenstates basis (see,
e.g., Ref.~\cite{EgoVol22}).

In Refs.~\cite{Dvo25,Dvo25b}, we extended the QFT formalism to account
for neutrino interactions with external fields. The dressed propagators
of neutrino mass eigentates, exactly taking into account the electroweak
interaction with matter and an external magnetic field were derived.
Note that we have obtained both diagonal and nondiagonal propagators.
The exact propagators are based on the solutions of the Dyson equations.
Alternative field theory approaches for the description of neutrino
oscillations in external fields were developed in Refs.~\cite{CarChu99,Dvo11,AkhWil13}

The results of Refs.~\cite{Dvo25,Dvo25b} are applied for Majorana
neutrinos. The direct application of the QFT formalism for the Dirac
particles is ambiguous. We mentioned above that the nature neutrinos
is unclear. In our work, we explore how QFT can be used to describe oscillations of Dirac neutrinos in external fields.

The present work is organized in the following way. After recalling
the basic issues of neutrino masses and mixing in Sec.~\ref{sec:NUMASSMIX},
we describe how QFT is applied to neutrino flavor oscillations in
external fields in Sec.~\ref{sec:QFTOSC}.

Then, in Sec.~\ref{sec:FLOSCMATT},
we study the situation of neutrinos propagating in background matter.
We remind how neutrinos can interact with matter within the standard
model in Sec.~\ref{subsec:NUMATTINT}. Then, we derive the dressed
propagators of Dirac neutrinos interacting with background matter
in Sec.~\ref{subsec:DRESSPROPMATT}. The matrix element and the transition
probability for oscillations of Dirac neutrinos in matter are obtained
in Sec.~\ref{subsec:QFTMSW}.

The impact of a magnetic field on the spin-flavor precession of Dirac
neutrinos in frames of QFT is considered in Sec.~\ref{sec:SFP}.
We briefly recall the basics of the electrodynamics of Dirac neutrinos
in Sec.~\ref{subsec:NUELECTROD}. The QM description of the spin-flavor
precession of Dirac neutrinos having a transition magnetic moment
is described in Sec.~\ref{sec:QMSFP}. The dressed propagators of
Dirac neutrinos in a magnetic field are derived in Sec.~\ref{subsec:DRESSPROPB}.
The matrix element and the corresponding probabilities of the spin-flavor
precession of Dirac neutrinos in frames of QFT are obtained in Sec.~\ref{subsec:QFTSFP}.

We consider some general problems which arise in the application of QFT to neutrino oscillations in Sec.~\ref{sec:GENISS}.
Finally, we conclude in Sec.~\ref{sec:CONCL}. The diagonal propagators of Dirac neutrinos in matter and in vacuum
are derived in Appendices~\ref{sec:DIRPROPMATT} and~\ref{sec:DIRPROPVAC}.
We provide some details for the computation of the matrix elements
in Appendix~\ref{sec:COMPINT}.

\section{Neutrino masses and mixing}\label{sec:NUMASSMIX}

The Lagrangian of the standard model is formulated in terms of active
flavor neutrinos, which interact with other fermions. However, these
particles do not have definite masses. In general situation, the mass
matrix has both left and right Majorana terms, as well as the Dirac
term. Thus, after the diagonalization of the mass matrix of the general
form, one gets $3+N_{s}$ mass eigenstates which turn out to be Majorana
particles with different masses. Here, $N_{s}\geq0$ is the number
of sterile neutrinos which are not excluded experimentally.

In this work, we adopt a specific situation when only a Dirac mass
term is present, i.e. the corresponding term in the Lagrangian
reads
\begin{equation}\label{eq:massterm}
  \mathcal{L}_{m}=
  -\sum_{\lambda\lambda'}m_{\lambda\lambda'}\bar{\nu}_{\lambda\mathrm{L}}\nu_{\lambda'\mathrm{R}}
  +\mathrm{h.c.},
\end{equation}
where $(m_{\lambda\lambda'})$ is the nondiagonal mass matrix. Moreover,
we consider a simplified case of two active flavor neutrinos, i.e.
$\nu_{\lambda}=(\nu_{e},\nu_{\mu})$. In this situation, the matrix
transformation, required to diagonalize the mass term in Eq.~(\ref{eq:massterm}),
has the form,
\begin{equation}\label{eq:flmassrel}
  \nu_{\lambda}=\sum_{a}U_{\lambda a}\psi_{a},\quad(U_{\lambda a})=
  \left(
    \begin{array}{cc}
      \cos\theta & \sin\theta\\
      -\sin\theta & \cos\theta
    \end{array}
  \right),
\end{equation}
where $\theta$ is the vacuum mixing angle. 

The total vacuum Lagrangian, rewritten in terms of the mass eigenstates
$\psi_{a}$, $a=1,2$, reads
\begin{equation}\label{eq:massLagr}
  \mathcal{L}=\sum_{a}\bar{\psi}_{a}(\mathrm{i}\gamma^{\mu}\partial_{\mu}-m_{a})\psi_{a},
\end{equation}
where $\psi_{a}$ are the Dirac fields, i.e. $\psi_{a}^{c}\neq\psi_{a}$,
having masses $m_{a}$, and $\gamma^{\mu}=(\gamma^{0},\bm{\gamma})$
are the Dirac matrices.

\section{QFT treatment of neutrino oscillations}\label{sec:QFTOSC}

As one see in Eq.~(\ref{eq:massLagr}), neutrino mass eigenstates
propagate in space as free particles. However, neutrino flavor eigenstates
convert one to another because of the mixing in Eq.~(\ref{eq:flmassrel}).
This process is called neutrino flavor oscillations. As a rule, neutrino
flavor oscillations are described in terms of the QM approach. However,
we mentioned in Sec.~\ref{sec:INTR} that the QM treatment reveals
certain shortcomings. Thus, the QFT approach for neutrino oscillations
has to be developed.

\begin{figure}
  \centering
  \includegraphics{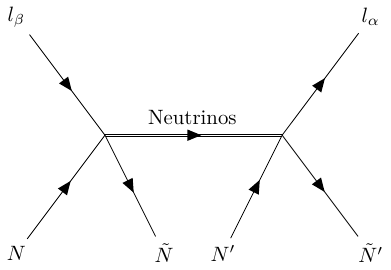}
  \protect
  \caption{The schematic illustration of the process corresponding to the $S$-matrix
element in Eq.~(\ref{eq:Smatr}). The broad neutrino line means that
these particles interact with an external field while propagating
from a source to a detector.\label{fig:diagschem}}
\end{figure}

We follow the QFT formalism proposed in Refs.~\cite{Kob82,GiuKimLee93,GriSto96}
to study neutrino flavor oscillations in vacuum. Neutrinos are treated
as virtual particles in this approach. We assume that a charged lepton
$l_{\beta}$ interacts with a neutrino source, which is a heavy nucleus.
A neutrino beam is created in the wake of this interaction, $N+l_{\beta}\to\tilde{N}+\text{neutrinos}$.
Then, the neutrino beam is absorbed by a detector, which can be again
a heavy nucleus. A charged lepton $l_{\alpha}$ is emitted by the
nucleus in the detector, $N'+\text{neutrinos}\to\tilde{N}'+l_{\alpha}$.
If $l_{\alpha}\neq l_{\beta}$, one can say that neutrino oscillations
happen while particles propagate from a source to a detector. We
show this kind of process schematically in Fig.~\ref{fig:diagschem}.

To study the aforementioned flavor transformations in the leptonic
sector, which are interpreted as neutrino oscillations, one has to
analyze the following $S$-matrix element:
\begin{equation}\label{eq:Smatr}
  S=-\frac{1}{2}
  \left(
    \sqrt{2}G_{\mathrm{int}}
  \right)^{2}
  \int\mathrm{d}^{4}x\mathrm{d}^{4}y
  \left\langle
    \tilde{N},\tilde{N}',l_{\alpha}
    \left|
      T
      \left\{
        j_{\mu}^{\dagger}(x)J^{\mu}(x)j^{\nu}(y)J_{\nu}^{\dagger}(y)
      \right\}
    \right|
    N,N',l_{\beta}
  \right\rangle,
\end{equation}
where
\begin{equation}\label{eq:lepcurr}
  j_{\mu}=\sum_{\lambda}\bar{\nu}_{\lambda\mathrm{L}}\gamma_{\mu}l_{\lambda\mathrm{L}},
\end{equation}
is operator valued the leptonic current, $J^{\mu}$ is the nuclear
current operator, and $G_{\mathrm{int}}$ is the coupling constant.
We assume that the neutrino emission and detection is treated in frames
of the standard model. That is why only left-handed chiral projections
of leptons are involved in Eq.~(\ref{eq:lepcurr}). For example,
$l_{\lambda\mathrm{L}}=\tfrac{1}{2}(1-\gamma^{5})l_{\lambda}$, where
$\gamma^{5}=\mathrm{i}\gamma^{0}\gamma^{1}\gamma^{2}\gamma^{3}$,
etc.

If nuclei $N$ and $\tilde{N}$, as well as $N'$ and $\tilde{N}'$,
are heavy, the averaging of $J^{\mu}$ over nuclear Fock states reads
\begin{equation}\label{eq:nuclcurr}
  \left\langle
    \tilde{N}'
    \left|
      J_{\mu}(x_{0},\mathbf{x})
    \right|
    N'
  \right\rangle
  \propto
  \delta_{\mu0}\delta(\mathbf{x}-\mathbf{x}_{2}),
  \quad
  \left\langle
    \tilde{N}
    \left|
      J_{\nu}^{\dagger}(y_{0},\mathbf{y})
    \right|
    N
  \right\rangle
  \propto
  \delta_{\nu0}\delta(\mathbf{y}-\mathbf{x}_{1}),
\end{equation}
where $\mathbf{x}_{1}$ and $\mathbf{x}_{2}$ are the positions of
a source and a detector. Then, we decompose the flavor neutrinos in
the mass basis in Eq.~(\ref{eq:flmassrel}) and assume that incoming
and outgoing charged leptons are plane waves having the momenta $p_{\beta}^{\mu}=(E_{\beta},\mathbf{p}_{\beta})$
and $p_{\alpha}^{\mu}=(E_{\alpha},\mathbf{p}_{\alpha})$. In this
case,
\begin{align}\label{eq:lepav}
  \left\langle
    l_{\alpha}
    \left|
      T
      \left\{
        j_{\mu}^{\dagger}(x)j_{\nu}(y)
      \right\}
    \right|
    l_{\beta}
  \right\rangle = &
  e^{\mathrm{i}p_{\alpha}x-\mathrm{i}p_{\beta}y}\bar{u}_{\alpha}(p_{\alpha})\gamma_{\mu}^{\mathrm{L}}
  \notag
  \\
  & \times
  \sum_{ab}U_{\alpha a}U_{\beta b}^{*}
  \left\langle
    0
    \left|
      T
      \left\{
        \psi_{a}(x)\bar{\psi}_{b}(y)
      \right\}
    \right|
    0
  \right\rangle
  \gamma_{\nu}^{\mathrm{L}}u_{\beta}(p_{\beta}),
\end{align}
where $\gamma_{\mu}^{\mathrm{L}} = \gamma_{\mu} (1-\gamma^5)/2$.

Finally, using Eqs.~(\ref{eq:nuclcurr}) and~(\ref{eq:lepav}),
Eq.~(\ref{eq:Smatr}) can be rewritten as
\begin{equation}\label{eq:EaEb}
  S=-2\pi G_{\mathrm{int}}^{2}\delta(E_{\alpha}-E_{\beta})
  e^{-i\mathbf{p}_{\alpha}\mathbf{x}_{2}+i\mathbf{p}_{\beta}\mathbf{x}_{1}}
  \mathrm{i}\mathcal{M}_{\beta\to\alpha},
\end{equation}
where
\begin{equation}\label{eq:matrelgen}
  \mathcal{M}_{\beta\to\alpha}=
  \bar{u}_{\alpha}(p_{\alpha})\gamma_{0}^{\mathrm{L}}
  \sum_{ab}U_{\alpha a}U_{\beta b}^{*}
  \left(
    \int\frac{\mathrm{d}^{3}q}{(2\pi)^{3}}e^{\mathrm{i}\mathbf{qL}}\Sigma_{ab}(E,\mathbf{q})
  \right)
  \gamma_{0}^{\mathrm{L}}u_{\beta}(p_{\beta}),
\end{equation}
is the matrix element, $E=(E_{\alpha}+E_{\beta})/2$ is the mean energy
of incoming and outgoing leptons, $\mathbf{L}=\mathbf{x}_{2}-\mathbf{x}_{1}$
is the vector connecting a source and a detector, and $\Sigma_{ab}(q)$
is the Fourier image of the propagator of the mass eigenstates $\Sigma_{ab}(x-y)=-\mathrm{i}\left\langle 0\left|T\left\{ \psi_{a}(x)\bar{\psi}_{b}(y)\right\} \right|0\right\rangle $.

If neutrino mass eigenstates, which are virtual particles, propagate
in vacuum, the propagators are diagonal in neutrino types, $\Sigma_{ab}\propto\delta_{ab}$.
In this situation, one can reproduce the known expressions for the
probability of neutrino oscillations in vacuum, $P_{\nu_{\beta}\to\nu_{\alpha}}\propto|\mathcal{M}_{\beta\to\alpha}|^{2}$;
cf. Refs.~\cite{Kob82,GiuKimLee93,GriSto96}.

If we take the two neutrinos system, we can consider, e.g., $\nu_{e}\to\nu_{\mu}$
oscillations. In this case, accounting for Eq.~(\ref{eq:flmassrel}),
Eq.~(\ref{eq:matrelgen}) takes the form,
\begin{align}\label{eq:matrel2nu}
  \mathcal{M}_{e\to\mu} = &
  \bar{u}_{\mu}(p_{\mu})\gamma_{0}^{\mathrm{L}}\int\frac{\mathrm{d}^{3}q}{(2\pi)^{3}}e^{\mathrm{i}\mathbf{qL}}
  \notag
  \\
  & \times
  \left[
    \sin\theta\cos\theta(\Sigma_{22}-\Sigma_{11})+\cos^{2}\theta\Sigma_{21}-\sin^{2}\theta\Sigma_{12}
  \right]
  \gamma_{0}^{\mathrm{L}}u_{e}(p_{e}).
\end{align}
We shall see shortly in Secs.~\ref{subsec:DRESSPROPMATT} and~\ref{subsec:DRESSPROPB}
that $\Sigma_{ab}$ acquire nondiagonal entries with $a\neq b$ if
neutrinos interact with external fields while they propagate from
a source to a detector. Our main goal is to find these nondiagonal
propagators in external fields and compute the corresponding probabilities.

\section{Flavor oscillations of Dirac neutrinos in matter}\label{sec:FLOSCMATT}

In this section, we develop the application of QFT for the description
of neutrino flavor oscillations in background matter. First, we remind
how a Dirac neutrino can interact with background fermions via eletroweak
forces. Then, we apply QFT to study neutrino oscillations in this
background. In particular, we calculate the matrix element and the
transition probability.

\subsection{Neutrino interaction with matter}\label{subsec:NUMATTINT}

We adopt the standard model neutrino interaction with background fermions.
As in Sec.~\ref{sec:NUMASSMIX}, we suppose that we deal with two
neutrino flavors, $\nu_{e}$ and $\nu_{\mu}$. The Lagrangian of the
active flavor neutrinos, interacting with nonmoving and unpolarized
matter consisting of electrons, protons and neutrons, is
\begin{equation}\label{eq:Lintmatt}
  \mathcal{L}_{\mathrm{int}}=-\frac{1}{2}\sum_{\lambda}V_{\lambda}\bar{\nu}_{\lambda}\gamma^{0}(1-\gamma^{5})\nu_{\lambda},
\end{equation}
where
\begin{equation}\label{eq:VeVmu}
  V_{\nu_{e}}=\sqrt{2}G_{\mathrm{F}}
  \left(
    n_{e}-\frac{1}{2}n_{n}
  \right),
  \quad
  V_{\nu_{\mu}}=-\frac{G_{\mathrm{F}}}{\sqrt{2}}n_{n},
\end{equation}
are the effective potentials of the flavor neutrinos interaction with
matter, $G_{\mathrm{F}}=1.17\times10^{-5}\,\text{GeV}^{-2}$ is the
Fermi constant, $n_{e}$ and $n_{n}$ are the number densities of
electrons and neutrons. In Eq.~(\ref{eq:Lintmatt}), we imply the
forward scattering approximation. It is important that the interaction
with background matter is diagonal in the neutrino flavor basis in
Eq.~(\ref{eq:Lintmatt}). It is the feature of the standard model.

When we introduce the neutrino mass eigenstates in Eq.~(\ref{eq:flmassrel}),
as required by the formalism in Sec.~\ref{sec:QFTOSC}, the neutrino
matter interaction becomes
\begin{equation}\label{eq:Lintmattmass}
  \mathcal{L}_{\mathrm{int}}=-\frac{1}{2}\sum_{ab}g_{ab}\bar{\psi}_{a}\gamma^{0}(1-\gamma^{5})\psi_{b},
\end{equation}
where
\begin{equation}\label{eq:gmatr}
  (g_{ab})=
  \left(
    \begin{array}{cc}
      g_{1} & g\\
      g & g_{2}
    \end{array}
  \right)=
  \sum_{\lambda}V_{\lambda}U_{\lambda a}^{*}U_{\lambda b},
\end{equation}
is the matrix of the effective potentials of massive neutrinos interaction
with matter.

It should be noted that $(g_{ab})$ is Eq.~(\ref{eq:gmatr}) is nondiagonal
in a general situation, i.e. $g_{12}=g_{21}\equiv g\neq0$. It means
that the corresponding Dirac equations for $\psi_{a}$, resulting
from Eqs.~(\ref{eq:massLagr}) and~(\ref{eq:Lintmattmass}), are
coupled,
\begin{align}\label{eq:Direqmattmass}
  \left[
    \mathrm{i}\gamma^{\mu}\partial_{\mu}-m_{1}-\frac{g_{1}}{2}\gamma^{0}(1-\gamma^{5})
  \right]
  \psi_{1} & =
  \frac{g}{2}\gamma^{0}(1-\gamma^{5})\psi_{2},
  \nonumber
  \\
  \left[
    \mathrm{i}\gamma^{\mu}\partial_{\mu}-m_{2}-\frac{g_{2}}{2}\gamma^{0}(1-\gamma^{5})
  \right]
  \psi_{2} & =
  \frac{g}{2}\gamma^{0}(1-\gamma^{5})\psi_{1}.
\end{align}
Equation~(\ref{eq:Direqmattmass}) means that the neutrino mass eigenstates
are converted into one another in background matter. As we see shortly
in Sec.~(\ref{subsec:DRESSPROPMATT}), it leads to the dressed propagators
$\Sigma_{ab}$, accounting for the matter interaction, nondiagonal
in neutrino types. That is, both $\Sigma_{12}$ and $\Sigma_{21}$
are nonzero.

\subsection{Dressed propagators of Dirac neutrinos in matter}\label{subsec:DRESSPROPMATT}

The major challenge in the exact accounting for matter effects in
the propagators of mass eigenstates is the nondiagonal potential $G=g\gamma^{0}(1-\gamma^{5})/2$
which mixes different mass states; cf. Eq.~(\ref{eq:Direqmattmass}).
The diagonal neutrino interaction with matter can be taken into account
in the propagators by using the exact solutions of the corresponding
Dirac equation with $g_{a}\neq0$ (see Appendix~\ref{sec:DIRPROPMATT}).

\begin{figure}[htbp]
  \centering
  \subfigure[]
  {\label{fig:f2a}
  \includegraphics[scale=0.75]{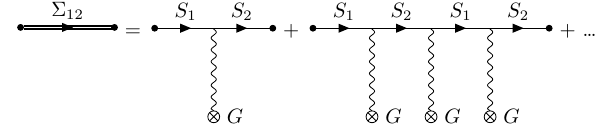}}
  \subfigure[]
  {\label{fig:f2b}
  \includegraphics[scale=0.75]{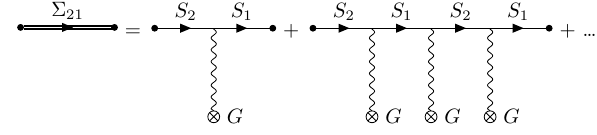}}
  \\
  \subfigure[]
  {\label{fig:f2c}
  \includegraphics[scale=0.75]{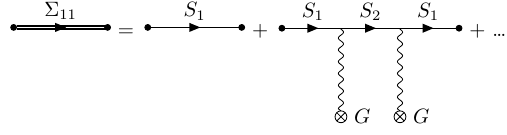}}
  \subfigure[]
  {\label{fig:f2d}
  \includegraphics[scale=0.75]{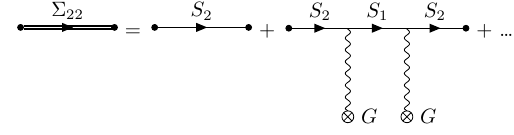}}
  \protect 
\caption{The Feynman diagrams contributing to the dressed propagators $\Sigma_{ab}$
of Dirac mass eigenstates which take into account both diagonal and
nondiagonal interactions with background matter. Thin lines correspond
to the propagators $S_{a}$ in Eq.~(\ref{eq:propdiagur}) in which
only diagonal interaction with matter is accounted for. The same diagrams
represent $\Sigma_{ab}$ for Dirac neutrinos with a transition magnetic
moment in a magnetic field. In a magnetic case, we replace $G\to V_{\mathrm{B}}=\mu(\bm{\Sigma}\mathbf{B})$.
Then, thin lines stand for the vacuum propagators in Eq.~(\ref{eq:Dirpropvac}).\label{fig:feyndiag}}
\end{figure}

We elaborated the formalism for finding $\Sigma_{ab}$ in the presence
of nonzero $G$ in Ref.\ \cite{Dvo25}. It is based on the summation
of the infinite series of Feynman diagrams shown in Fig.~\ref{fig:feyndiag}.
Each thin line corresponds to a propagator $S_{a}$ which accounts
for the diagonal matter potential $g_{a}$. Such propagators are given
in Eq.~(\ref{eq:propdiagur}). The external field is given by the
mixing matter potential $G$.

The result of the summation of series in Fig.~\ref{fig:feyndiag}
can be represented as the Dyson equations for $\Sigma_{ab}$,
\begin{align}
  (S_{1}GS_{2})^{-1}-G & =\Sigma_{12}^{-1},
  \quad
  (S_{2}GS_{1})^{-1}-G=\Sigma_{21}^{-1},
  \label{eq:Dyseqab}
  \\
  S_{1}^{-1}-GS_{2}G & =\Sigma_{11}^{-1},
  \quad
  S_{2}^{-1}-GS_{1}G=\Sigma_{22}^{-1}.
  \label{eq:Dyseqaa}
\end{align}
The details of the derivation of Eqs.~(\ref{eq:Dyseqab}) and~(\ref{eq:Dyseqaa})
are given in Ref.\ \cite{Dvo25}. In Ref.\ \cite{Dvo25}, we solved
Eqs.~(\ref{eq:Dyseqab}) and~(\ref{eq:Dyseqaa}) for Majorana neutrinos
for which the potential $G$ is the $c$-number.

For Dirac neutrinos the matrix $G$ is singular since it contains
the projection operator, $G\propto(1-\gamma^{5})/2$. Hence, a straightforward
solution of Eq.~(\ref{eq:Dyseqab}) is ambiguous since it involves
the reciprocal of $G$. Nevertheless, we can regularize this potential
by replacing it with
\begin{equation}\label{eq:Greg}
  G\to\frac{g}{2}\gamma^{0}(1-\alpha\gamma^{5}).
\end{equation}
Here $\alpha\to1$ in the final expression for $\Sigma_{ab}$. We
also regularize the diagonal propagators $S_{a}$ by introducing the
factors $\xi<1$ and $\alpha_{a}<1$, which both are equal to one
for ultrarelativistic neutrinos, to avoid the divergences in getting
$S_{a}^{-1}$ in Eqs.~(\ref{eq:Dyseqab}) and~(\ref{eq:Dyseqaa}).
We set $\xi\to1$ and $\alpha_{a}\to1$ after the calculation of $\Sigma_{ab}$
is made.

Let us provide the detailed computations of, e.g., $\Sigma_{12}$.
Based on Eqs.~(\ref{eq:Greg}) and~(\ref{eq:propdiagur}), we get
that
\begin{align}\label{eq:GSarecipr}
  G^{-1} & =\frac{2}{g}\frac{1+\alpha\gamma^{5}}{1-\alpha^{2}}\gamma^{0},
  \nonumber
  \\
  S_{a}^{-1} & =-\frac{4(p_{0}-E_{a-}-g_{a}/2+\mathrm{i}0)}{(1-\alpha_{a}^{2})(1-\xi^{2})}
  \left(
    \begin{array}{cc}
      0 & 1+\xi(\bm{\sigma}\hat{\mathbf{p}})\\
      1+\xi(\bm{\sigma}\hat{\mathbf{p}}) & 0
    \end{array}
  \right)(1+\alpha_{a}\gamma^{5}),
\end{align}
where we use the notation $\hat{\mathbf{p}} = \mathbf{p}/p$ for a unit vector along $\mathbf{p}$.

Based on Eq.~(\ref{eq:Dyseqab}) and using Eq.~(\ref{eq:GSarecipr}),
one gets that
\begin{equation}\label{eq:Sigma12recip}
  \Sigma_{12}^{-1}=
  \left(
    \begin{array}{cc}
      0 & D_{1}\\
      D_{2} & 0
    \end{array}
  \right),
\end{equation}
where
\begin{align}\label{eq:Sigma12cont}
  D_{1} & =-A(a-b)C+\frac{g}{2}(1+\alpha),
  \nonumber
  \\
  D_{2} & =-A(a+b)C+\frac{g}{2}(1-\alpha),
  \nonumber
  \\
  A & =\frac{32(p_{0}-E_{1-}-g_{1}/2+\mathrm{i}0)(p_{0}-E_{2-}-g_{2}/2+\mathrm{i}0)}
  {g(1-\alpha^{2})(1-\alpha_{1}^{2})(1-\alpha_{2}^{2})(1-\xi^{2})^{2}},
  \nonumber
  \\
  C & =1+\xi^{2}+2\xi(\bm{\sigma}\hat{\mathbf{p}}),
  \nonumber
  \\
  a & =1+\alpha(\alpha_{1}+\alpha_{2})+\alpha_{1}\alpha_{2},
  \nonumber
  \\
  b & =\alpha+\alpha_{1}+\alpha_{2}+\alpha\alpha_{1}\alpha_{2}.
\end{align}
Inverting the matrix in Eq.~(\ref{eq:Sigma12recip}) and using Eq.~(\ref{eq:Sigma12cont}),
one obtains that
\begin{equation}\label{eq:Sigma12arbE}
  \Sigma_{12}=
  \left(
    \begin{array}{cc}
      0 & D_{2}^{-1}\\
      D_{1}^{-1} & 0
    \end{array}
  \right),
\end{equation}
where
\begin{align}\label{eq:D12recip}
  D_{1}^{-1} & =\frac{g(1+\alpha)/2-A(a-b)(1+\xi^{2})+2A\xi(a-b)(\bm{\sigma}\hat{\mathbf{p}})}{
  \left[
    g(1+\alpha)/2-A(a-b)(1+\xi^{2})
  \right]^{2}
  -4A^{2}\xi^{2}(a-b)^{2}},
  \nonumber
  \\
  D_{2}^{-1} & =\frac{g(1-\alpha)/2-A(a+b)(1+\xi^{2})+2A\xi(a+b)(\bm{\sigma}\hat{\mathbf{p}})}{
  \left[
    g(1-\alpha)/2-A(a+b)(1+\xi^{2})
  \right]^{2}
  -4A^{2}\xi^{2}(a+b)^{2}}.
\end{align}
For ultrarelativistic neutrinos we should set $\alpha_{1,2}\to1$.
Moreover, to remove the regularization in the propagators $S_{a}$
and in the potential $G$, we approach the limits $\xi\to1$ and $\alpha\to1$.
As a result, we get that
\begin{equation}
  D_{1}^{-1}\to-\frac{g[1-(\bm{\sigma}\hat{\mathbf{p}})]}{2[(p_{0}-E_{1-}-g_{1}/2+\mathrm{i}0)(p_{0}-E_{2-}-g_{2}/2+\mathrm{i}0)-g^{2}]},
\end{equation}
and $D_{2}^{-1}\to0$. The final expression for $\Sigma_{12}$ reads
\begin{equation}\label{eq:Sigma12}
  \Sigma_{12}=-\frac{g}{2[(p_{0}-E_{1-}-g_{1}/2+\mathrm{i}0)(p_{0}-E_{2-}-g_{2}/2+\mathrm{i}0)-g^{2}]}
  \left(
    \begin{array}{cc}
      0 & 0\\
      1-(\bm{\sigma}\hat{\mathbf{p}}) & 0
    \end{array}
  \right),
\end{equation}
where $E_{1,2-}$ are given in Eq.~(\ref{eq:enlevmatt}), with $\sigma=-1$.
Making similar calculations, one gets that $\Sigma_{21}=\Sigma_{12}$.

Analogously, one can find the expressions for $\Sigma_{aa}$. We present
them in the final form for ultrarelativistic neutrinos after all regularizations
are removed,
\begin{align}\label{eq:Sigmaaa}
  \Sigma_{11} & =-\frac{p_{0}-E_{2-}-g_{2}/2}{2[(p_{0}-E_{1-}-g_{1}/2+\mathrm{i}0)(p_{0}-E_{2-}-g_{2}/2+\mathrm{i}0)-g^{2}]}
  \left(
    \begin{array}{cc}
      0 & 0\\
      1-(\bm{\sigma}\hat{\mathbf{p}}) & 0
    \end{array}
  \right),
  \nonumber
  \\
  \Sigma_{22} & =-\frac{p_{0}-E_{1-}-g_{1}/2}{2[(p_{0}-E_{1-}-g_{1}/2+\mathrm{i}0)(p_{0}-E_{2-}-g_{2}/2+\mathrm{i}0)-g^{2}]}
  \left(
    \begin{array}{cc}
      0 & 0\\
      1-(\bm{\sigma}\hat{\mathbf{p}}) & 0
    \end{array}
  \right).
\end{align}
Note that, if one turns off the nondiagonal matter interaction in
Eq.~(\ref{eq:Sigmaaa}), $g\to0$, one gets that $\Sigma_{aa}\to S_{a}$,
as it should be. Indeed, if we set $\alpha_{a}\to1$ and $\xi\to1$
in Eq.~(\ref{eq:propdiagur}), we obtain that
\begin{equation}
  S_{a}\to-\frac{1}{2(p_{0}-E_{a-}-g_{a}/2+\mathrm{i}0)} 
  \left(
    \begin{array}{cc}
      0 & 0\\
      1-(\bm{\sigma}\hat{\mathbf{p}}) & 0
    \end{array}
  \right).
\end{equation}
That is, our calculations are correct in the limiting case.

\subsection{Matrix element and transition probability in matter}\label{subsec:QFTMSW}

To proceed with the calculation of the matrix element in Eq.~(\ref{eq:matrel2nu})
we choose the coordinate system so that $\mathbf{L}=L\mathbf{e}_{z}$,
with $L>0$. Then, for simplicity, we assume the forward scattering
approximation for charged leptons. In this case, $u_{e,\mu}^\mathrm{T}=(0,0,0,1)$. Thus, both the incoming
electron and the outgoing muon are left particles propagating along
the $z$-axis. Moreover, we rewrite the energies of left neutrinos
as $E_{a-}\approx E_{a}+g_{a}/2$, where $E_{a}=\sqrt{q^{2}+m_{a}^{2}}$
is the energy of a free particle. We also use the cylindrical coordinates
$(\rho,z,\phi)$ for the momentum of a virtual neutrino, $\mathbf{q}=\rho\mathbf{e}_{\rho}+z\mathbf{e}_{z}$.

Using Eqs.~(\ref{eq:Sigma12}) and~(\ref{eq:Sigmaaa}), we rewrite
Eq.~(\ref{eq:matrel2nu}) as
\begin{align}\label{eq:matrelcylcoord}
  \mathcal{M}_{e\to\mu} = &
  \frac{1}{16\pi^{2}}\int_{0}^{\infty}\rho\mathrm{d}\rho\int_{-\infty}^{+\infty}\mathrm{d}z
  \left(
    1+\frac{z}{\sqrt{z^{2}+\rho^{2}}}
  \right)
  e^{\mathrm{i}zL}
  \notag
  \\
  & \times
  \frac{\frac{1}{2}(E_{2}-E_{1}+g_{2}-g_{1})\sin2\theta+g\cos2\theta}{\sqrt{
  \left(
    \frac{E_{2}-E_{1}+g_{2}-g_{1}}{2}
  \right)^{2}
  +g^{2}}}
  \left(
    \frac{1}{E-\mathcal{E}_{+}+\mathrm{i}0}-\frac{1}{E-\mathcal{E}_{-}+\mathrm{i}0}
  \right),
\end{align}
where
\begin{equation}
  \mathcal{E}_{\pm}=\frac{1}{2}(E_{2}+E_{1}+g_{2}+g_{1})\pm\sqrt{
  \left(
    \frac{E_{2}-E_{1}+g_{2}-g_{1}}{2}
  \right)^{2}
  +g^{2}}.
\end{equation}
The integration in Eq.~(\ref{eq:matrelcylcoord}) is carried out
similarly to that in Refs.~\cite{Dvo25,Dvo25b}. Some details of
the integrals computation can be found in Appendix~\ref{sec:COMPINT}.
We just present the matrix element in the final form,
\begin{equation}\label{eq:Memug}
  \mathcal{M}_{e\to\mu} \approx
  -\frac{\mathrm{i}\bar{E}_{m}e^{\mathrm{i}\bar{E}_{m}L}}{2\pi L\mathfrak{E}_{m}}
  \sin(\mathfrak{E}_{m}L)
  \left[
    \left(
      \frac{\Delta m^{2}}{4E}+\frac{g_{2}-g_{1}}{2}
    \right)
    \sin2\theta+g\cos2\theta
  \right]
  \left(
    1+\frac{\mathrm{i}}{2L\bar{E}_m}
  \right),
\end{equation}
where
\begin{equation}\label{eq:Em}
  \bar{E}_{m}=E+\frac{m_{2}^{2}+m_{1}^{2}}{4E}+\frac{g_{2}+g_{1}}{2},
  \quad
  \mathfrak{E}_{m}=\sqrt{\left(\frac{\Delta m^{2}}{4E}+\frac{g_{2}-g_{1}}{2}\right)^{2}+g^{2}},
\end{equation}
and $\Delta m^{2} = m_2^2 - m_1^2 > 0$. We neglect terms $\sim (\mathfrak{E}_{m}/\bar{E}_{m})^2 \ll 1$ in deriving Eq.~\eqref{eq:Memug}.

One finds the effective potentials of the
mass eigenstates interaction with matter in Eqs.~(\ref{eq:Memug})
and~(\ref{eq:Em}),
\begin{equation}\label{eq:g1g2g}
g_{1}=V_{\nu_{e}}\cos^{2}\theta+V_{\nu_{\mu}}\sin^{2}\theta,\quad g_{2}=V_{\nu_{e}}\sin^{2}\theta+V_{\nu_{\mu}}\cos^{2}\theta,\quad g=(V_{\nu_{e}}-V_{\nu_{\mu}})\sin\theta\cos\theta,
\end{equation}
with help of Eqs.~(\ref{eq:flmassrel}) and~(\ref{eq:gmatr}). Taking into account Eqs.~(\ref{eq:VeVmu}) and~(\ref{eq:Memug})-(\ref{eq:g1g2g}),
one gets the transition probability for flavor oscillations of Dirac
neutrinos in matter, $P_{\nu_{e}\to\nu_{\mu}} \propto |\mathcal{M}_{e\to\mu}|^{2}$.

In squaring of the matrix element, we omit the term $E^2_m/4\pi^2 \approx E^2/4\pi^2$ since it is the common multiplier in the transition probability. In practice, this factor can be eliminated by the proper normalization of the interaction rate of incoming charged leptons with a nucleus in a neutrino source. Additionally, one has the factor $1/L^2$ which the transition probability is proportional to. It stays to provide the inverse square law which the flux of neutrinos, emitted in a source, obeys. Since this factor is the same for all neutrino types, i.e. it does not influence the dynamics of neutrino oscillations, we also omit it.

Eventually, we get that the total transition probability can be represented in the following form: $P_{\nu_{e}\to\nu_{\mu}} = P_{\nu_{e}\to\nu_{\mu}}^{(\text{qm})} + P_{\nu_{e}\to\nu_{\mu}}^{(\text{qft})}$
where
\begin{align}\label{eq:PMSW}
  P_{\nu_{e}\to\nu_{\mu}}^{(\text{qm})} = &
  \frac{
  \left(
    \frac{\Delta m^{2}}{4E}\sin2\theta
  \right)^{2}}{
  \left(
    \frac{\Delta m^{2}}{4E} \sin2\theta
  \right)^{2}+
  \left(
    \frac{\Delta m^{2}}{4E}\cos2\theta-\frac{G_{\mathrm{F}}n_{e}}{\sqrt{2}}
  \right)^{2}}
  \notag
  \\
  & \times
  \sin^{2}
  \left(
    \sqrt{
    \left(
      \frac{\Delta m^{2}}{4E}\sin2\theta
    \right)^{2}+
    \left(
      \frac{\Delta m^{2}}{4E}\cos2\theta-\frac{G_{\mathrm{F}}n_{e}}{\sqrt{2}}
    \right)^{2}}L
  \right),
  \notag
  \\
  P_{\nu_{e}\to\nu_{\mu}}^{(\text{qft})} = & P_{\nu_{e}\to\nu_{\mu}}^{(\text{qm})}/4L^2E^2.
\end{align}
In Eq.~\eqref{eq:PMSW}, we keep the terms up to $L^{-2}$.

Formally, $P_{\nu_{e}\to\nu_{\mu}}^{(\text{qm})}$ in Eq.~(\ref{eq:PMSW}) coincides with the prediction of the
QM approach for neutrino oscillations in matter~\cite{Wol78,MikSmi85}.
However, in QM, the quantity $E$ is referred to a mean energy of
a neutrino beam, $E_{\nu}$, which is a not well defined parameter
if one deals with a system of flavor neutrinos. We recall that, in
the QFT treatment, $E=(E_{\alpha}+E_{\beta})/2$. Thus, this quantity
is related to the energies of charged leptons, which are the observables
since these particles correspond to in and out states.

The correction to the transition probability, given by $P_{\nu_{e}\to\nu_{\mu}}^{(\text{qft})}$, results from the QFT treatment of neutrino oscillations. This quantity is suppressed when a sufficiently long propagation distance is considered. However, if $L$ is macroscopic but relatively short, then $P_{\nu_{e}\to\nu_{\mu}}^{(\text{qft})}$ can contribute to the total probability.

Analogous corrections to the probability were studied in Ref.~\cite{NauShk21}, where it was suggested that this kind of contributions can resolve the reactor antineutrino anomaly (see, e.g., Ref.~\cite{ZhaQiaFal24}). Note that the computed correction to the transition probability turns out to be positive,  $P_{\nu_{e}\to\nu_{\mu}}^{(\text{qft})} = \delta >0$. Hence, the total survival probability is suppressed by the same amount, $P_{\nu_{e}\to\nu_{e}} = P_{\nu_{e}\to\nu_{e}}^{(\text{qm})} - \delta$. Therefore, our result is consistent with Ref.~\cite{NauShk21}.

\section{Spin-flavor precession of Dirac neutrinos in a magnetic field}\label{sec:SFP}

In this section, we apply QFT to study the spin-flavor precession
of Dirac neutrinos in magnetic fields. We start with the basics of
the neutrino electrodynamics. Then, we recall how to describe the
neutrino spin-flavor precession in frames of the QM approach. Finally,
we use QFT to reproduce the results of QM and study the corrections
to the QM transition probability.

\subsection{Interaction of Dirac neutrinos with an electromagnetic field}\label{subsec:NUELECTROD}

Despite neutrinos are electrically neutral particles, i.e. their electric
charges are equal to zero~\cite{millichnu}, nothing prevents them
to have magnetic moments. Neutrino magnetic moments are defined for
neutrino mass eigenstates. The neutrino electromagnetic interaction
is described by the following Lagrangian,
\begin{equation}\label{eq:Lintmag}
\mathcal{L}_{\mathrm{int}}=-\frac{1}{2}\sum_{ab}\mu_{ab}\bar{\psi}_{a}\sigma^{\mu\nu}\psi_{b}F_{\mu\nu},
\end{equation}
where $\sigma^{\mu\nu}=\tfrac{\mathrm{i}}{2}[\gamma^{\mu},\gamma^{\nu}]_{-}$
are the Dirac matrices, $F_{\mu\nu}=(\mathbf{E},\mathbf{B})$ is the
electromagnetic field tensor, $\mathbf{E}$ and $\mathbf{B}$ are
the electric and magnetic field strengths.

The matrix of magnetic moments $(\mu_{ab})$ in Eq.~(\ref{eq:Lintmag})
is Hermitian for Dirac neutrinos~\cite{Pas00}. The diagonal and
off-diagonal entries of this matrix are called the diagonal and transition
magnetic moments, respectively. Unlike charged leptons, neutrino magnetic
moments have purely anomalous nature and are caused by the neutrino
interaction with the vacuum of the underlying model. In our analysis,
we assume that transition magnetic moments dominate over the diagonal
ones. In some models of neutrino magnetic moments, one has the opposite
situation because of the Glashow--Iliopoulos--Maiani mechanism~\cite{GlaIliMai70}.
Nevertheless, one can discuss arbitrary values of magnetic moments
from the phenomenological point of view. Our choice in favor of the
transition magnetic moment is dictated by the fact that we would like
to highlight the external field which is essentially nondiagonal in
neutrino types.

The magnetic moments for neutrino flavor eigenstates can be formally
represented in the form,
\begin{equation}
M_{\lambda\lambda'}=\sum_{ab}\mu_{ab}U_{a\lambda}U_{b\lambda'}^{*},
\end{equation}
where we use the fact that the mixing matrix in Eq.~(\ref{eq:flmassrel}) is unitary.

If only the magnetic field is present, $\mathbf{E}=0$ and $\mathbf{B}\neq0$,
we shall see shortly in Sec.~\ref{sec:QMSFP} that the transitions
between neutrinos belonging to different flavors and opposite helicity
states are possible provided that these neutrinos have a transition
magnetic moment. This process is called the spin-flavor
precession.

\subsection{QM description of the neutrino spin-flavor precession}\label{sec:QMSFP}

As in Sec.~\ref{sec:FLOSCMATT}, we study two flavor neutrinos, $\nu_{e}$
and $\nu_{\mu}$. As a rule, neutrino oscillations are described in
the flavor basis in frames of the QM approach. However, owing to the
fact that both the nature of neutrinos and the neutrino magnetic moments
are defined for neutrino mass eigenstates, we adopt the mass basis
here. The corresponding mass eigenstates are supposed to be Dirac
particles. We take that these massive neutrinos have only the transition
magnetic moment $\mu$ and interact with a magnetic field $\mathbf{B}=(B,0,0)$
transverse with respect to the neutrino momentum which is along the
$z$-axis.

We use the basis $\Psi^{\mathrm{T}}=(\psi_{1\mathrm{L}},\psi_{2\mathrm{L}},\psi_{1\mathrm{R}},\psi_{2\mathrm{R}})$
in the QM description. The effective Schr\"odinger equation for this
system reads
\begin{equation}\label{eq:effScheq}
  \mathrm{i}\frac{\mathrm{d}\Psi}{\mathrm{d}t}=H\Psi,
  \quad
  H=
  \left(
    \begin{array}{cccc}
      \frac{m_{1}^{2}}{2E_{\nu}} & 0 & 0 & -\mu B\\
      0 & \frac{m_{2}^{2}}{2E_{\nu}} & -\mu B & 0\\
      0 & -\mu B & \frac{m_{1}^{2}}{2E_{\nu}} & 0\\
      -\mu B & 0 & 0 & \frac{m_{2}^{2}}{2E_{\nu}}
    \end{array}
  \right),
\end{equation}
where $E_{\nu}$ is the mean neutrino energy. We omit the term proportional
to the unit matrix in the effective Hamiltonian $H$ in Eq.~(\ref{eq:effScheq}).

The general solution of Eq.~(\ref{eq:effScheq}) has the form,
\begin{equation}\label{eq:GensolScheq}
  \Psi(t)=
  \left[
    \left(
      U_{1}\otimes U_{1}^{\dagger}+U_{2}\otimes U_{2}^{\dagger}
    \right)
    e^{-iE_{\nu}^{(+)}t}+
    \left(
      V_{1}\otimes V_{1}^{\dagger}+V_{2}\otimes V_{2}^{\dagger}
    \right)
    e^{-iE_{\nu}^{(-)}t}
  \right]
  \Psi_{0},
\end{equation}
where
\begin{align}
  U_{1} & =\frac{1}{\sqrt{2\mathfrak{E}_{\nu}}}
  \left(
    \begin{array}{c}
      \sqrt{\mathfrak{E}_{\nu}-\frac{\Delta m^{2}}{4E_{\nu}}}\\
      0\\
      0\\
      -\frac{\mu B}{\sqrt{\mathfrak{E}_{\nu}-\frac{\Delta m^{2}}{4E_{\nu}}}}
    \end{array}
  \right),
  \quad
  U_{2}=\frac{1}{\sqrt{2\mathfrak{E}_{\nu}}}
  \left(
    \begin{array}{c}
      0\\
      \sqrt{\mathfrak{E}_{\nu}+\frac{\Delta m^{2}}{4E_{\nu}}}\\
      -\frac{\mu B}{\sqrt{\mathfrak{E}_{\nu}+\frac{\Delta m^{2}}{4E_{\nu}}}}\\
      0
    \end{array}
  \right),
  \nonumber
  \\
  V_{1} & =\frac{1}{\sqrt{2\mathfrak{E}_{\nu}}}
  \left(
    \begin{array}{c}
      \sqrt{\mathfrak{E}_{\nu}+\frac{\Delta m^{2}}{4E_{\nu}}}\\
      0\\
      0\\
      \frac{\mu B}{\sqrt{\mathfrak{E}_{\nu}+\frac{\Delta m^{2}}{4E_{\nu}}}}
    \end{array}
  \right),
  \quad
  V_{2}=\frac{1}{\sqrt{2\mathfrak{E}_{\nu}}}
  \left(
    \begin{array}{c}
      0\\
      \sqrt{\mathfrak{E}_{\nu}-\frac{\Delta m^{2}}{4E_{\nu}}}\\
      \frac{\mu B}{\sqrt{\mathfrak{E}_{\nu}-\frac{\Delta m^{2}}{4E_{\nu}}}}\\
      0
    \end{array}
  \right),
\end{align}
are the eigenvectors and
\begin{equation}
  E_{\nu}^{(\pm)}=\frac{m_{1}^{2}+m_{2}^{2}}{4E_{\nu}}\pm\mathfrak{E}_{\nu},
  \quad
  \mathfrak{E}_{\nu}=\sqrt{
  \left(
    \frac{\Delta m^{2}}{4E_{\nu}}
  \right)
  +(\mu B)^{2}},
\end{equation}
are the eigenvalues of $H$.

The initial condition $\Psi_{0}$ in Eq.~(\ref{eq:GensolScheq})
is taken as $\Psi_{0}^{\mathrm{T}}=(\cos\theta,\sin\theta,0,0)$.
It is based on Eq.~(\ref{eq:flmassrel}) and the fact that initially
we have only left-handed electron neutrinos, $N^{\mathrm{T}}(0)=(1,0,0,0)$,
where $N^{\mathrm{T}}=(\nu_{e\mathrm{L}},\nu_{\mu\mathrm{L}},\nu_{e\mathrm{R}},\nu_{\mu\mathrm{R}})$
is the effective wavefunction of flavor neutrinos. Now, Eq.~(\ref{eq:GensolScheq})
takes the form,
\begin{equation}\label{eq:Psigensol}
  \Psi(t)=\frac{1}{\mathfrak{E}_{\nu}}
  \left(
    \begin{array}{c}
      \left[
        \mathfrak{E}_{\nu}\cos\mathfrak{E}_{\nu}t+\mathrm{i}\frac{\Delta m^{2}}{4E_{\nu}}\sin\mathfrak{E}_{\nu}t
      \right]
      \cos\theta\\
      \left[
        \mathfrak{E}_{\nu}\cos\mathfrak{E}_{\nu}t-\mathrm{i}\frac{\Delta m^{2}}{4E_{\nu}}\sin\mathfrak{E}_{\nu}t
      \right]
      \sin\theta\\
      \mathrm{i}\mu B\sin\theta\sin\mathfrak{E}_{\nu}t\\
      \mathrm{i}\mu B\cos\theta\sin\mathfrak{E}_{\nu}t
    \end{array}
  \right),
\end{equation}
which obeys the initial condition.

Using Eqs.~(\ref{eq:flmassrel}) and~(\ref{eq:Psigensol}), we get
the time dependent flavor neutrinos wavefunction,
\begin{equation}\label{eq:Nugensol}
  N(t)=
  \left(
    \begin{array}{c}
      \cos\mathfrak{E}_{\nu}t+\mathrm{i}\frac{\Delta m^{2}}{4\mathfrak{E}_{\nu}E_{\nu}}\cos2\theta\sin\mathfrak{E}_{\nu}t\\
      -\mathrm{i}\frac{\Delta m^{2}}{4\mathfrak{E}_{\nu}E_{\nu}}\sin2\theta\sin\mathfrak{E}_{\nu}t\\
      \mathrm{i}\frac{\mu B}{\mathfrak{E}_{\nu}}\sin2\theta\sin\mathfrak{E}_{\nu}t\\
      \mathrm{i}\frac{\mu B}{\mathfrak{E}_{\nu}}\cos2\theta\sin\mathfrak{E}_{\nu}t
    \end{array}
  \right).
\end{equation}
Based on Eq.~(\ref{eq:Nugensol}), we derive the probabilities for
the whole range of transformations in a neutrino beam,
\begin{align}
  P_{\nu_{e\mathrm{L}} \to\nu_{e\mathrm{L}}}(t) & =
  \cos^{2}\left(\mathfrak{E}_{\nu}t\right)+
  \left(
    \frac{\Delta m^{2}}{4\mathfrak{E}_{\nu}E_{\nu}}
  \right)^{2}
  \cos^{2}(2\theta)\sin^{2}
  \left(
    \mathfrak{E}_{\nu}t
  \right),
  \label{eq:PeLeL}
  \\
  P_{\nu_{e\mathrm{L}}\to\nu_{\mu\mathrm{L}}}(t) & =
  \left(
    \frac{\Delta m^{2}}{4\mathfrak{E}_{\nu}E_{\nu}}
  \right)^{2}
  \sin^{2}(2\theta)\sin^{2}
  \left(
    \mathfrak{E}_{\nu}t
  \right),
  \label{eq:PeLmuL}
  \\
  P_{\nu_{e\mathrm{L}}\to\nu_{e\mathrm{R}}}(t) & =
  \left(
    \frac{\mu B}{\mathfrak{E}_{\nu}}
  \right)^{2}
  \sin^{2}(2\theta)\sin^{2}
  \left(
    \mathfrak{E}_{\nu}t
  \right),
  \label{eq:PeLeR}
  \\
  P_{\nu_{e\mathrm{L}}\to\nu_{\mu\mathrm{R}}}(t) & =
  \left(
    \frac{\mu B}{\mathfrak{E}_{\nu}}
  \right)^{2}
  \cos^{2}(2\theta)\sin^{2}
  \left(
    \mathfrak{E}_{\nu}t
  \right).
  \label{eq:PeLmuR}
\end{align}
The sum of the probabilities in Eqs.~(\ref{eq:PeLeL})-(\ref{eq:PeLmuR})
equals to one.

\subsection{Dressed propagators of Dirac neutrinos in magnetic field}\label{subsec:DRESSPROPB}

The main technique for finding of the dressed propagators of Dirac
neutrinos in a magnetic field is analogous to that in Sec.~\ref{subsec:DRESSPROPMATT}.
The sequence of Feynman diagrams contributing to $\Sigma_{ab}$ is
shown in Fig.~\ref{fig:feyndiag}. The explanation why the diagrams
for $\Sigma_{ab}$ in a magnetic field formally coincide with those
in the matter case will be given shortly in Sec.~\ref{subsec:QFTSFP}.
We just mention that one does not need to consider the propagators
of left and right neutrinos before and after each interaction with
a magnetic field, as it was made for Majorana neutrinos in Ref.~\cite{Dvo25b}.
Thin lines in Fig.~\ref{fig:feyndiag} correspond to Eq.~(\ref{eq:Dirpropvac}),
with the correct chiral projection being chosen automatically. We
recall that we consider the situation when only a transition magnetic
moment is present in the system of two Dirac neutrinos.

The Dyson equations for $\Sigma_{ab}$ in a magnetic field coincide
with those in Eqs.~(\ref{eq:Dyseqab}) and~(\ref{eq:Dyseqaa}) with
the replacement $G\to\mu(\bm{\Sigma}\mathbf{B})$, where $\bm{\Sigma}=\gamma^{5}\gamma^{0}\bm{\gamma}$.
Since one has to invert $S_{a}$ in Eqs.~(\ref{eq:Dyseqab}) and~(\ref{eq:Dyseqaa}),
we regularize it by inserting the factor $\varsigma_{a}<1$ in Eq.~(\ref{eq:Dirpropvac})
to avoid singularities in $S_{a}^{-1}$. One should set $\varsigma_{a}\to1$
after a final expression for $\Sigma_{ab}$ is obtained.

Let us provide some details, e.g., for the $\Sigma_{11}$ calculations.
After tedious but straightforward computations, the solution of Eq.~(\ref{eq:Dyseqaa})
for $\Sigma_{11}$ takes the form,
\begin{equation}\label{eq:Sigma11B}
  \Sigma_{11}=\frac{\gamma^{0}l^{0}-(\bm{\gamma}\mathbf{l})}{l_{0}^{2}-l^{2}},
\end{equation}
where
\begin{align}\label{eq:lol}
  l_{0} & =\frac{2(p_{0}-E_{1}+\mathrm{i}0)}{1-\varsigma_{1}^{2}}
  -\frac{\mu^{2}B^{2}}{2(p_{0}-E_{2}+\mathrm{i}0)},
  \nonumber
  \\
  \mathbf{l} & =\hat{\mathbf{p}}
  \left(
    \frac{2\varsigma_{1}(p_{0}-E_{1}+\mathrm{i}0)}{1-\varsigma_{1}^{2}}
    +\frac{\mu^{2}B^{2}\varsigma_{2}}{2(p_{0}-E_{2}+\mathrm{i}0)}
  \right)-
  \frac{\mu^{2}\varsigma_{2}(\mathbf{B}\hat{\mathbf{p}})\mathbf{B}}
  {(p_{0}-E_{2}+\mathrm{i}0)}.
\end{align}
Removing the regularization in Eqs.~(\ref{eq:Sigma11B}) and~(\ref{eq:lol})
by setting $\varsigma_{1,2}\to1$, one gets that
\begin{equation}\label{eq:Sigma11Bfin}
  \Sigma_{11}=\frac{(p_{0}-E_{2})(\gamma^{0}-\bm{\gamma}\hat{\mathbf{p}})}
  {2[(p_{0}-E_{1}+\mathrm{i}0)(p_{0}-E_{2}+\mathrm{i}0)-\mu^{2}B^{2}+\mu^{2}(\mathbf{B}\hat{\mathbf{p}})^{2}]}.
\end{equation}
Analogously to Eq.~(\ref{eq:Sigma11Bfin}), we obtain that
\begin{equation}\label{eq:Sigma22Bfin}
  \Sigma_{22}=\frac{(p_{0}-E_{1})(\gamma^{0}-\bm{\gamma}\hat{\mathbf{p}})}
  {2[(p_{0}-E_{1}+\mathrm{i}0)(p_{0}-E_{2}+\mathrm{i}0)-\mu^{2}B^{2}+\mu^{2}(\mathbf{B}\hat{\mathbf{p}})^{2}]}.
\end{equation}
One can see in Eqs.~(\ref{eq:Sigma11Bfin}) and~(\ref{eq:Sigma22Bfin})
that, at $B\to0$, $\Sigma_{aa}\to S_{a}$, where $S_{a}$ is given
in Eq.~(\ref{eq:Dirpropvac}), as it should be.

We shall see shortly in Sec.~\ref{subsec:QFTSFP} that $\Sigma_{12}$
and $\Sigma_{21}$, resulting from Eq.~(\ref{eq:Dyseqab}), do not
contribute to the process which we identify with the spin-flavor precession.
That is why we do not present these propagators here for the sake
of brevity.

The denominators of $\Sigma_{aa}$ in Eqs.~(\ref{eq:Sigma11Bfin})
and~(\ref{eq:Sigma22Bfin}) contain the term $\propto\mu^{2}(\mathbf{B}\cdot\hat{\mathbf{p}})^{2}$.
Analogous terms are present in the dressed propagators of Majorana
neutrinos in a magnetic field, which were obtained in Ref.~\cite{Dvo25b}.
In the QM description of the spin-flavor precession in Sec.~\ref{sec:QMSFP},
the magnetic field is taken to be transverse to the neutrino momentum.
Thus, in the QM approach, one is allowed to set $(\mathbf{B}\cdot\hat{\mathbf{p}})=0$ since neutrinos are on the mass
shell. In QFT treatment, we cannot drop the term
$\propto\mu^{2}(\mathbf{B}\cdot\hat{\mathbf{p}})^{2}\neq0$ in Eqs.~(\ref{eq:Sigma11Bfin})
and~(\ref{eq:Sigma22Bfin}) since $\mathbf{p}$ is the momentum of
a virtual neutrino. In fact, $\mathbf{p}$ can be arbitrary since
we integrate over it in Eq.~(\ref{eq:matrel2nu}). Note that the
quantity in question, $\propto\mu^{2}(\mathbf{B}\cdot\hat{\mathbf{p}})^{2}$,
is an essentially quantum term in the propagator which arises from
the summation of the infinite number of the Feynman diagrams in Fig.~\ref{fig:feyndiag}.
Shortly in Sec.~\ref{subsec:QFTSFP}, we shall compute the contributions
to the matrix element and the transition probability originating from
this term.

\subsection{QFT description of the spin-flavor precession in a magnetic field}\label{subsec:QFTSFP}

We have derived all kinds of transition probabilities for spin-flavor
precession of Dirac neutrinos in Sec.~\ref{sec:QMSFP}. Despite the
transitions like $\nu_{e\mathrm{L}}\to\nu_{(e,\mu)\mathrm{R}}$ are
formally allowed, see Eqs.~(\ref{eq:PeLeR}) and~(\ref{eq:PeLmuR}),
we cannot detect right-handed Dirac neutrinos. We assume that a neutrino
detector is based on the standard model physics, i.e. it involves
the leptonic current in Eq.~(\ref{eq:lepcurr}). This current contains
only the left-handed neutrino fields. That is why the processes $\nu_{e\mathrm{L}}\to\nu_{\mu\mathrm{R}}$ and $\nu_{e\mathrm{L}}\to\nu_{e\mathrm{R}}$
cannot be considered in frames of the adopted model for the description of oscillations if one deals with ultrarelativistic neutrinos.

It is the key difference between the spin-flavor precession of Majorana
and Dirac neutrinos. We recall that the marker of the spin-flavor
precession in the Majorana case is the appearance of an anti-charged-lepton.
That is, if we consider the process $\nu_{e}\xrightarrow{B}\bar{\nu}_{\mu}$
for Majorana $\nu_{e}$ and $\nu_{\mu}$, one has an electron $e^{-}$
in a source and antimuon $\mu^{+}$ in a detector~\cite{Dvo25b}.
Therefore, to study the impact of the magnetic field on the spin-flavor
precession of Dirac neutrinos in frames of the QFT based approach,
one has to consider the transitions $\nu_{e\mathrm{L}}\to\nu_{(e,\mu)\mathrm{L}}$
and try to rederive Eqs.~(\ref{eq:PeLeL}) and~(\ref{eq:PeLmuL}).
Since, in the initial and final states, we deal with left particles only,
the Feynman diagrams in Fig.~\ref{fig:feyndiag} in the magnetic
and matter cases formally coincide.

We shall study, e.g., the process $\nu_{e\mathrm{L}}\to\nu_{\mu\mathrm{L}}$.
The matrix element for this oscillation channel is given by Eq.~(\ref{eq:matrel2nu}).
One can see in Figs.~\ref{fig:f2a} and~\ref{fig:f2b}
that the propagators $\Sigma_{12}$ and $\Sigma_{21}$, which arise
from Eq.~(\ref{eq:Dyseqab}), correspond to the spin-flip, i.e. the
transitions like $\psi_{1,2\mathrm{L}}\leftrightarrow\psi_{2,1\mathrm{R}}$.
Indeed, each interaction of mass eigenstates with $V_{\mathrm{B}}$
makes their spin to flip. Since the diagrams in Figs.~\ref{fig:f2a} and~\ref{fig:f2b} have odd number of interactions with the
nondiagonal external field, a spin-flip occurs in any order of the
perturbation theory. Thus, the propagators $\Sigma_{12}$ and $\Sigma_{21}$
do not contribute to $\mathcal{M}_{e\to\mu}$ since we should have
the same helicities of the initial and the final states.

In our calculations, we adopt the same geometry of the system as in
Sec.~\ref{subsec:QFTMSW}. Namely, $\mathbf{L}=L\mathbf{e}_{z}$.
Analogously to Sec.~\ref{sec:QMSFP}, we take that the magnetic field
is transverse to the line connecting the source and the detector,
$\mathbf{B}=B\mathbf{e}_{x}$. Then, we also assume the forward scattering
approximation for charged leptons, i.e., $u_{e,\mu}^\mathrm{T}=(0,0,0,1)$.

Finally, the matrix element reads,
\begin{equation}\label{eq:matrelSFPgen}
  \mathcal{M}_{e\to\mu}=\frac{\sin2\theta}{4}
  \int\frac{\mathrm{d}^{3}q}{(2\pi)^{3}}e^{\mathrm{i}\mathbf{qL}}
  \frac{(E_{2}-E_{1})(1+\hat{q}_{z})}{(E-E_{1}+\mathrm{i}0)(E-E_{2}+\mathrm{i}0)-\mu^{2}\tilde{B}^{2}},
\end{equation}
where $E_{1,2}=\sqrt{q^{2}+m_{1,2}^{2}}$ are the energies of free
mass eigenstates and $\tilde{B}^{2}=B^{2}(1-\hat{q}{}_{x}^{2})$ is
the effective magnetic field. Note that the quantity $\mu B^{2}\hat{q}{}_{x}^{2}$
originates from the quantum term $\propto\mu^{2}(\mathbf{B}\cdot\hat{\mathbf{p}})^{2}$
in the propagators in Eqs.~(\ref{eq:Sigma11Bfin}) and~(\ref{eq:Sigma22Bfin}).

The details of the computation of integrals in Eq.~(\ref{eq:matrelSFPgen})
are provided in Appendix~\ref{sec:COMPINT}. We just present the
final result as
\begin{equation}\label{eq:martelB}
  \mathcal{M}_{e\to\mu}=-\frac{\mathrm{i}Ee^{\mathrm{i}EL}}{2\pi L}
  \frac{\sin2\theta\frac{\Delta m^{2}}{4E}\sin\mathfrak{E}_{\mathrm{B}}L}{\mathfrak{E}_{\mathrm{B}}}
  \left(
    1-\frac{(\mu B)^{2}}{2E\mathfrak{E}_{\mathrm{B}}}
  \right)
  \left[  
    1 +\frac{\mathrm{i}}{2EL}
    \left(
      1-\frac{(\mu B)^{2}}{2\mathfrak{E}_\mathrm{B}^{2}}
    \right)
  \right],
\end{equation}
where
\begin{equation}\label{eq:EB}
  \mathfrak{E}_{\mathrm{B}}=\sqrt{\left(\frac{\Delta m^{2}}{4E}\right)+(\mu B)^{2}}.
\end{equation}
Analogously to Eq.~(\ref{eq:Memug}), Eq.~(\ref{eq:martelB}) is
valid for for ultrarelativistic neutrinos. We also keep the terms up to $L^{-2}$ in Eq.~\eqref{eq:martelB}.

The transition probability for the process $\nu_{e\mathrm{L}}\to\nu_{\mu\mathrm{L}}$ is
\begin{equation}\label{eq:PnueLnumuLQFT}
  P_{\nu_{e\mathrm{L}}\to\nu_{\mu\mathrm{L}}}
  \propto|\mathcal{M}_{e\to\mu}|^{2}
  \propto P_{\mathrm{max}}\sin^{2}
  \left(
    \sqrt{(\mu B)^{2}+
    \left(
      \frac{\Delta m^{2}}{4E}
    \right)^{2}}L
  \right),
\end{equation}
where $P_{\mathrm{max}}\approx P_{\nu_{e\mathrm{L}}\to\nu_{\mu\mathrm{L}}}^{(\mathrm{qm})}+P_{\nu_{e\mathrm{L}}\to\nu_{\mu\mathrm{L}}}^{\prime(\mathrm{qft})}+P_{\nu_{e\mathrm{L}}\to\nu_{\mu\mathrm{L}}}^{\prime\prime(\mathrm{qft})}$, and
\begin{align}
  P_{\nu_{e\mathrm{L}}\to\nu_{\mu\mathrm{L}}}^{(\mathrm{qm})} & =
  \sin^{2}2\theta\frac{
  \left(
    \frac{\Delta m^{2}}{4E}
  \right)^{2}}
  {(\mu B)^{2}+
  \left(
    \frac{\Delta m^{2}}{4E}
  \right)^{2}},
  \label{eq:PmaxQM}
  \\
  P_{\nu_{e\mathrm{L}}\to\nu_{\mu\mathrm{L}}}^{\prime(\mathrm{qft})} & =
  -\frac{(\mu B)^{2}}{E\mathfrak{E}_\mathrm{B}}P_{\nu_{e\mathrm{L}}\to\nu_{\mu\mathrm{L}}}^{(\mathrm{qm})},
  \label{eq:PmaxQFT}
  \\
  P_{\nu_{e\mathrm{L}}\to\nu_{\mu\mathrm{L}}}^{\prime\prime(\mathrm{qft})} & =
  \frac{1}{4E^2L^2}
  \left(
    1-\frac{(\mu B)^{2}}{2\mathfrak{E}_\mathrm{B}^2}
  \right)^{2}
  P_{\nu_{e\mathrm{L}}\to\nu_{\mu\mathrm{L}}}^{(\mathrm{qm})}.
  \label{eq:PmaxQFTL}
\end{align}
One can see that Eqs.~(\ref{eq:PnueLnumuLQFT}) and~(\ref{eq:PmaxQM})
reproduce the QM result in Eq.~(\ref{eq:PeLmuL}) if we identify
the mean neutrino energy $E_{\nu}$ with $E=(E_{e}+E_{\mu})/2$, as
well as assume that the time $t$ equals to the distance $L$ traveled
by a neutrino beam.

The correction to the transition probability $P_{\nu_{e\mathrm{L}}\to\nu_{\mu\mathrm{L}}}^{\prime(\mathrm{qft})}$
in Eq.~(\ref{eq:PmaxQFT}) arises from the term $\propto\mu^{2}(\mathbf{B}\cdot\hat{\mathbf{p}})^{2}$
in the propagators in Eqs.~(\ref{eq:Sigma11Bfin}) and~(\ref{eq:Sigma22Bfin}),
which is discussed at the end of Sec.~\ref{subsec:DRESSPROPB}. One
can see that $P_{\nu_{e\mathrm{L}}\to\nu_{\mu\mathrm{L}}}^{(\mathrm{qft})}<0$,
i.e. the total transition probability is less than the QM prediction.
Moreover, one gets from Eqs.~(\ref{eq:PmaxQM}) and~(\ref{eq:PmaxQFT})
that $|P_{\nu_{e\mathrm{L}}\to\nu_{\mu\mathrm{L}}}^{\prime(\mathrm{qft})}|/P_{\nu_{e\mathrm{L}}\to\nu_{\mu\mathrm{L}}}^{(\mathrm{qm})}\sim(\mu B)^{2}/\mathfrak{E}_{\mathrm{B}}E\sim\mu B/E\ll1$
provided that the magnetic contribution is dominant. Therefore, the
QFT correction to the propagators, $\propto\mu^{2}(\mathbf{B}\cdot\hat{\mathbf{p}})^{2}$,
can be safely neglected since its contribution to the transition probability
is negligible. We also mention that the QFT correction to the probability
of the spin-flavor precession of Majorana neutrinos, $\nu_{e}\xrightarrow{B}\bar{\nu}_{\mu}$,
analogous to that in Eq.~(\ref{eq:PmaxQFT}), was obtained in Ref.~\cite{Dvo25b}.

The QFT correction to the transition probability in Eq.~\eqref{eq:PmaxQFTL}, $P_{\nu_{e\mathrm{L}}\to\nu_{\mu\mathrm{L}}}^{\prime\prime(\mathrm{qft})}$ results from keeping the next-to-leading term $\propto L^{-2}$ in Eq.~\eqref{eq:martelB}. This contribution is analogous to the QFT correction derived in Sec.~\ref{subsec:QFTMSW}. One can see that $P_{\nu_{e\mathrm{L}}\to\nu_{\mu\mathrm{L}}}^{\prime\prime(\mathrm{qft})}/P_{\nu_{e\mathrm{L}}\to\nu_{\mu\mathrm{L}}}^{(\mathrm{qm})} = 1/16E^2L^2$ provided that the magnetic contribution is dominant, i.e. $\mathfrak{E}_\mathrm{B} \sim \mu B$. Thus, $P_{\nu_{e\mathrm{L}}\to\nu_{\mu\mathrm{L}}}^{\prime\prime(\mathrm{qft})}$ can become important if short macroscopic propagation distances are considered.

At the end of this section, we mention that, analogously to the $\nu_{e\mathrm{L}}\xrightarrow{B}\nu_{\mu\mathrm{L}}$
oscillations channel, one can treat $\nu_{e\mathrm{L}}\xrightarrow{B}\nu_{e\mathrm{L}}$
transitions. We just present the final expression for the corresponding
survival probability
\begin{equation}\label{eq:PnuLnuLQFT}
  P_{\nu_{e\mathrm{L}}\to\nu_{e\mathrm{L}}}\approx
  \left[
    1-\frac{(\mu B)^{2}}{E\mathfrak{E}_{\mathrm{B}}}
  \right]
  \left(
    \cos^{2}\mathfrak{E}_{\mathrm{B}}L+
    \left(
      \frac{\Delta m^{2}}{4\mathfrak{E}_{\mathrm{B}}E}
    \right)^{2}\sin^{2}\mathfrak{E}_{\mathrm{B}}L
  \right).
\end{equation}
One can see in Eq.~(\ref{eq:PnuLnuLQFT}) that the leading term reproduces
the QM result in Eq.~(\ref{eq:PeLeL}). A small quantum correction
$\sim\mu B/E$ to the amplitude of the survival probability is also
present Eq.~(\ref{eq:PnuLnuLQFT}). In Eq.~\eqref{eq:PnuLnuLQFT}, we omit the QFT correction $\propto L^{-2}$. It can be derived analogously to Eq.~\eqref{eq:PmaxQFTL}.

\section{General issues of the QFT description of neutrino oscillations\label{sec:GENISS}}

Instead of dealing with nondiagonal propagators of neutrino mass eigenstates in external fields, one may consider the `propagators' of neutrino flavor states $S_\mathrm{F}$. These `propagators' are $4N_\nu \times 4N_\nu$ matrices, where $N_\nu$ is the number of the neutrino flavors. In vacuum, $S_\mathrm{F}$ obeys the modified Dirac equation (see, e.g., Ref.~\cite{KovSim24}),
\begin{equation}\label{eq:flpropdef1}
  (\mathrm{i}\gamma^\mu \partial_\mu - M)S_\mathrm{F}(x) = \delta^4(x),
\end{equation}
where $M$ is the nondiagonal mass matrix; cf. Eq.~\eqref{eq:massterm}. Equation~\eqref{eq:flpropdef1} is solved in the Fourier representation. The Fourier image of  $S_\mathrm{F}$ is singular on the mass shell. This singularity is removed by the proper poles bypassing in the standard manner. It is claimed that the poles bypassing gives the meaning of causality to the formal solution of Eq.~\eqref{eq:flpropdef1}.

As we demonstrated in Sec.~\ref{sec:QFTOSC}, the probability of neutrino oscillations is based on the element of the $S$-matrix,
\begin{equation}\label{eq:Smatrgen}
  S = \langle \Psi_\mathrm{out} | T\exp
  \bigg[
    \mathrm{i} \int \mathrm{d}^4 x \mathcal{L}_\mathrm{int}(x)
  \bigg]
  | \Psi_\mathrm{in} \rangle,
\end{equation}
where $\mathcal{L}_\mathrm{int}$ is the interaction Lagrangian and $| \Psi_\mathrm{in,out} \rangle$ are the in- and out- Fock states of particles represented as external lines in Fig.~\ref{fig:diagschem}. In the perturbative approach, the $S$-matrix element corresponds to the Feynman diagram shown in Fig.~\ref{fig:diagschem}, with neutrinos being represented by the internal line. If one adopts the concept of flavor neutrinos $\nu_\lambda$ as fundamental particles, one gets the following two points correlators while decomposing the $S$-matrix element:
\begin{equation}\label{eq:flpropdef2}
  \langle 0_\mathrm{F} | \nu_\lambda(x) \bar{\nu}_{\lambda'}(y) | 0_\mathrm{F} \rangle = (\tilde{S}_{\mathrm{F}})_{\lambda\lambda'}(x-y).
\end{equation}
by using the standard QFT formalism. In Eq.~\eqref{eq:flpropdef2}, $| 0_\mathrm{F} \rangle$ is the `vacuum' state of flavor neutrinos.

However, it is not obvious that Eq.~\eqref{eq:flpropdef2}, which appears in the calculation of probabilities, coincides with the solution of Eq.~\eqref{eq:flpropdef1} for a nondiagonal mass matrix $M$: $(\tilde{S}_{\mathrm{F}})_{\lambda\lambda'} \overset{?}{=} (S_{\mathrm{F}})_{\lambda\lambda'}$. The coincidence of two representations for the causal Green function, or the propagator, can be proven only for mass rather than flavor  neutrino eigenstates.
That is why, using the solution of Eq.~\eqref{eq:flpropdef1}, and its analogue in the presence of external fields~\cite{CarChu99,AkhWil13}, for the calculation of the probabilities is ambiguous.

There is one additional hidden problem with Eq.~\eqref{eq:flpropdef2}. Fock states of an \emph{elementary particle} correspond to irreducible representations of the Poincar\'e group~\cite{Bog90}. If a particle is massive, each representation is characterized by eigenvalue of the Casimir operator $P^\mu P_\mu = m_a^2$. Flavor neutrinos are the superposition mass states, cf. Eq.~\eqref{eq:flmassrel}. In this sense, all Fock states, including the vacuum one $| 0_\mathrm{F} \rangle$, of flavor neutrinos are not well defined from the point of view of QFT. Only if all masses of mass states are zero, $m_a = 0$, the vacua of flavor and mass states are unitary equivalent. Thus, according to the Weinberg definition~\cite{Wei95}, one can say that a flavor neutrino is not an \emph{elementary particle}.

On the contrary, the approach proposed in Refs.~\cite{Kob82,GiuKimLee93,GriSto96} is based on the neutrino mass eigenstates. Fock states and propagators of such particles are well defined in QFT. In the present work, we generalize the formalism of Refs.~\cite{Kob82,GiuKimLee93,GriSto96} to account for external fields. For this purpose, one has to deal with the nontrivial infinite sum of Feynman diagrams in Fig.~\ref{fig:feyndiag}. Nevertheless, in our analysis, we  do not go beyond well established methods of QFT.

We use the model of the neutrino emission, propagation and oscillations proposed in Ref.~\cite{Kob82}. In this approach, one makes two assumptions: (i) the distance between a source and a detector is constant, and (ii) the incoming and outgoing charged leptons propagate as plane waves, i.e. they have definite momenta. The assumption (i) results in equal energies of charged leptons in Eq.~\eqref{eq:EaEb}: $E_\alpha = E_\beta$. Since masses of these leptons are different, $\mathfrak{m}_\alpha \neq \mathfrak{m}_\beta$, their momenta momenta are not equal $\mathbf{p}_\alpha \neq \mathbf{p}_\beta$. To overcome the momentum nonconservation problem, we assume that charged leptons are ultrarelativistic in the present work.

In practice, heavy nuclei in a source and a detector are not at rest. Thus, instead of using the coordinate delta functions in Eq.~\eqref{eq:nuclcurr}, one should integrate over the positions a source and a detector, with the distance between them being kept constant. This procedure was shown in Ref.~\cite{Vol17} to be equivalent to the modification of the Feynman propagator. In this case, charged leptons still can be considered as plane waves.

Another possibility is the consideration of the wave packets of charged leptons. That is, $\mathbf{p}_{\alpha,\beta}$ are supposed to be spread around a certain mean momenta $\mathbf{p}_{\alpha,\beta}^{(0)}$. This approach was first proposed in Ref.~\cite{GiuKimLee93} and, then, developed in numerous works (see, e.g., Ref.~\cite{Beu03} for a review). The wave packet formalism allows to solve the energy-momentum conservation problem in neutrino oscillations.

In general, the violation of the above assumptions, either (i) or (ii), leads to the loss of coherence in neutrino oscillations. That is, the oscillatory behavior of the probabilities is suppressed when $L$ exceeds certain value called the coherence length $L_\mathrm{coh}$ (see, e.g., Refs.~\cite{GiuKimLee93,EgoVol19}). The computation of $L_\mathrm{coh}$ is one of the challenges to a QFT based approach for neutrino oscillations. In a realistic situation, numerous factors, such as the finite lifetime of charged leptons, the energy spread because of measurements etc., contribute to $L_\mathrm{coh}$. The recent calculation of $L_\mathrm{coh}$ was carried out in Ref.~\cite{DobMelSch25}. Note that the value of $L_\mathrm{coh}$ can depend on an experimental setup where neutrino oscillations are studied~\cite{GriStoMoh99}.

The majority of calculations of  $L_\mathrm{coh}$ are made for neutrino oscillations in vacuum. In the vacuum case, even if either the assumption (i) or (ii) is violated, one can still proceed with analytical calculations. When one deals with neutrino oscillations in external fields, the analytical calculation are difficult even in an idealized situation when both assumptions (i) and (ii) are valid (see Appendix~\ref{sec:COMPINT}). That is why, the study of decoherence in the QFT approach for neutrino oscillations in external fields, developed in the present work, is rather challenging.

\section{Discussion}\label{sec:CONCL}

In this work, we have applied QFT for the description of oscillations
of Dirac neutrinos in external fields. We have started in Sec.~\ref{sec:NUMASSMIX}
with the basics of the neutrino masses and mixing. Then, in Sec.~\ref{sec:QFTOSC},
we have recalled how oscillations can be treated in frames of the
QFT based approach, in which neutrinos are virtual particles. While
considering oscillations in vacuum, this formalism is valid for both
Majorana and Dirac neutrinos. However, the Dirac case has certain
peculiarities, compared to Majorana particles, when oscillations in
external fields are discussed. These features are analyzed in the
present work.

We have considered two cases of external fields. First, in Sec.~\ref{sec:FLOSCMATT},
we have studied flavor oscillations in matter supposing that neutrinos
are Dirac particles. We have reminded how neutrinos can interact with
background matter in the standard model in Sec.~\ref{subsec:NUMATTINT}.
Then, the dressed propagators of Dirac mass eigenstates in matter
have been derived in Sec.~\ref{subsec:DRESSPROPMATT}. The matrix
element and the transition probability for flavor oscillations have
been obtained in Sec.~\ref{subsec:QFTMSW}.

First, we mention that the main feature of the QFT approach, adopted
here, is utilizing the propagators of neutrino mass eigenstates. If
one considers neutrino oscillations in external fields, the exact
propagators turn out to be nondiagonal in neutrino types. It is because
of the fact that the interaction of neutrinos with external fields
mixes different mass eigenstates. Moreover, the poles structure of
these propagators acquires nontrivial contributions from the external
fields.

The technique for finding of the exact propagators is solving the
system of the Dyson equations, which are equivalent to summing of infinite
series of Feynman diagrams, depicted in Fig.~\ref{fig:feyndiag}.
In these diagrams, the nondiagonal interaction with matter is accounted
for. This technique was elaborated in Ref.~\cite{Dvo25} where oscillations
of Majorana neutrinos in matter were studied. In case of Dirac neutrinos,
the effective matter potential is a singular operator. Since the reciprocal
of this operator is involved in the Dyson equations, the straightforward
solution of these equations is ambiguous. Nevertheless, we have regularized
this potential in Sec.~\ref{subsec:DRESSPROPMATT}. After removing
the regularization, we have obtained the dressed propagators in background
matter in Eqs.~(\ref{eq:Sigma12}) and~(\ref{eq:Sigmaaa}). These
propagators have been used in the matrix element to get the transition
probability of flavor oscillations in matter in Eq.~(\ref{eq:PMSW}). The leading term in Eq.~(\ref{eq:PMSW}), $P_{\nu_{e}\to\nu_{\mu}}^{(\text{qm})}$, coincides with the prediction of the QM approach.

Then, we have studied the spin-flavor precession of Dirac neutrinos
in a magnetic field in Sec.~\ref{sec:SFP}. We have adopted the situation
when one has only a transition magnetic moment in the mass basis of
two neutrinos. This kind of magnetic moments is disfavored in some
models. Nevertheless, such an external magnetic interaction maximally mixes
different neutrino mass states. Thus, the QFT formalism for neutrino
oscillations based on dressed propagators is highlighted. Moreover,
one can consider arbitrary magnetic moments from the phenomenological
point of view.

We also mention that the impact of diagonal magnetic moments on the spin-flavor precession of Dirac neutrinos in frames of QFT was previously studied in Ref.~\cite{EgoVol22}. In this situation, it is not required to consider the Dyson Eqs.~\eqref{eq:Dyseqab} and~\eqref{eq:Dyseqaa} to get nondiagonal dressed propagators. The contribution of the magnetic field can be accounted for by using the exact solution of the Dirac equation for a single neutrino mass eigenstate. The corresponding propagator can be derived analogously to Appendix~\ref{sec:DIRPROPMATT}, where matter case was studied. On the contrary, if one deals with Dirac neutrinos with a transition magnetic moment, the application of the technique developed in the present work is essential to obtain the dressed propagators in a magnetic field. This fact was the additional motivation to consider this particular type of magnetic moments.

It would be interesting to consider the spin-flavor precession of Dirac neutrinos with an arbitrary matrix of magnetic moments in frames of QFT. However, the Feynman diagrams in Fig.~\ref{fig:feyndiag}, leading to the Dyson Eqs.~\eqref{eq:Dyseqab} and~\eqref{eq:Dyseqaa}, seem to be branching in this situation. The summation of this kind of infinite series is likely to be quite challenging. We plan to study this issue in one of our forthcoming works.

We have revealed a peculiarity of the treatment of the spin-flavor
precession of Dirac neutrinos in frames of QFT. This formalism involves
both a source, a neutrino propagation, and a detector as inherent
parts. Since our model of a detector is based on the standard model
physics, where right-handed neutrinos are sterile, we cannot describe
the channel like $\nu_{e\mathrm{L}}\to\nu_{\mu\mathrm{R}}$. On the contrary, the process like
$\nu_{e\mathrm{L}}\to\nu_{\mu\mathrm{R}}$ can be formally studied in QM description
which is provided in Sec.~\ref{sec:QMSFP}; cf. Eq.~(\ref{eq:PeLmuR}). In frames of the used model of neutrino oscillations, the spin-flavor precession of active left electron neutrinos is manifest as the deficit of these active particles.

It is the main difference between the Dirac and Majorana cases. In
the latter situation, the process like $\nu_{e}\to\bar{\nu}_{\mu}$
is well defined in the QFT formalism. The marker of such a spin-flavor
precession is the appearance of an antimuon in a detector. 

In frames of QFT, the impact of a magnetic field on the spin-flavor
precession of Dirac neutrinos can be judged indirectly by studying
the channels like $\nu_{e\mathrm{L}}\to\nu_{e\mathrm{L}}$ or $\nu_{e\mathrm{L}}\to\nu_{\mu\mathrm{L}}$,
which involve only left-handed particles. We have derived in details
the propagators contributing to the $\nu_{e\mathrm{L}}\to\nu_{\mu\mathrm{L}}$
channel. As in the matter case, these propagators are also based on
the solution of the Dyson equations which are equivalent to summing
up the infinite series of Feynman diagrams in Figs.~\ref{fig:f2c} and~\ref{fig:f2d}.

One can see in the propagators in Eqs.~(\ref{eq:Sigma11Bfin}) and~(\ref{eq:Sigma22Bfin})
that there is a quantum term $\propto\mu^{2}(\mathbf{B}\cdot\hat{\mathbf{p}})^{2}$
which cannot be neglected if we are in frames of QFT. We have derived
the matrix element and the transition probability based on these propagators
in Sec.~\ref{subsec:QFTSFP}. The leading term in the transition
probability in Eq.~(\ref{eq:PmaxQM}) reproduces the QM result in
Eq.~(\ref{eq:PeLmuL}). The correction to the QM expression is given
in Eq.~(\ref{eq:PmaxQFT}). It arises from the quantum term $\propto\mu^{2}(\mathbf{B}\cdot\hat{\mathbf{p}})^{2}$
in the propagators. This correction turns out to be small for ultrarelativistic
neutrinos for all reasonable parameters. Thus, we have revealed a
posteriori, i.e. after the calculation of a transition probability,
that the virtuality of massive neutrinos in propagators is negligible.
Note that the correction to the transition probability, analogous
to that in Eq.~(\ref{eq:PmaxQFT}), was derived in Ref.~\cite{Dvo25b}
while applying QFT for the description of the spin-flavor precesion
of Majorana neutrinos.

Besides the correction to the transition probabilities of the spin-flavor precession resulting from the QFT contributions to the propagators, we have derived the corrections arising from consideration of the short macroscopic propagation distance. These contributions are $\sim (EL)^{-2}$. The terms in question appear both in the transition probabilities of flavor oscillations in matter and the spin-flavor precession.

When one applies QFT to describe neutrino oscillations, we encounter to some general problems. They include the definition of the Fock states and the propagators of flavor neutrinos, as well as the decoherence in neutrino oscillations. These problems arise even if one studies neutrino oscillations in vacuum. We have discussed these issues in Sec.~\ref{sec:GENISS}.

We also mention that we have derived the diagonal propagators of massive
Dirac neutrinos in matter in Appendix~\ref{sec:DIRPROPMATT}, as
well as in vacuum in Appendix~\ref{sec:DIRPROPVAC}. Some details
of the computation of integrals, which one deals with in matrix elements,
have been provided in Appendix~\ref{sec:COMPINT}.

We also note that the impact of external fields on neutrino oscillations
is an essentially nonperturbative phenomenon which requires the summation
of infinite number of Feynman diagrams Fig.~\ref{fig:feyndiag}.
Only in this case, one gets the correct form of dressed propagators
exactly accounting for an external field, which result in the transition
probabilities consistent with the QM predictions.

We also list the main assumptions made in the
course of applying of QFT for neutrino oscillations in external fields.
The ultrarelativity of neutrinos, the consideration of two mass eigenstates,
and the spatial homogeneity of external fields were important in our analysis.

One of the most important assumption made in our work is the ultrarelativity of neutrinos. From the formal point of view, we neglect several terms in the undressed propagators (compare Eq.~\eqref{eq:propdiagmatt} with Eq.~\eqref{eq:propdiagur}, as well as Eq.~\eqref{eq:Dirproptransf} with Eq.~\eqref{eq:Dirpropvac}). Recently, in Ref.~\cite{Dvo26} we revealed the physical meaning of this kind of procedure. It was demonstrated to be equivalent to disregarding of the antiparticle degrees of freedom. This procedure of the truncation of the undressed propagators allows one to obtain the proper poles in the dressed propagators accounting for the two neutrino states. Note that the truncation of $S_a$ can be justified for ultrarelativistic neutrinos only.

However, if one considers the truncated undressed propagators $S_a$ in Eqs.~\eqref{eq:propdiagur} and Eq.~\eqref{eq:Dirpropvac}, as well as omits the terms $\sim m_a/E_a$ in the numerators of the integrands, $S_a$ become singular. While solving the Dyson Eqs.~\eqref{eq:Dyseqab} and~\eqref{eq:Dyseqaa}, one has to deal with the reciprocal of $S_a$ in the intermediate computations. That is, one, first, obtains $S_a^{-1}$. Then, this quantity is used to get the reciprocal of the dressed propagators, $\Sigma_{ab}^{-1}$. Finally, based on $\Sigma_{ab}^{-1}$, we derive $\Sigma_{ab}$.

Thus, one can guess that the singularity in $S_a$ does not affect the final form of $\Sigma_{ab}$ since we invert matrices two times. Nevertheless, certain regularization of $S_a$ is required. The same argument is applied to the regularization of the nondiagonal matter potential $G$ in Eq.~\eqref{eq:Greg}. Indeed, one can see in Eq.~\eqref{eq:GSarecipr} that both $G^{-1}$ and $S_a^{-1}$ are infinite when $\alpha = \alpha_a = \xi = 1$.

One can regularize of the undressed propagators in multiple ways. For example, we could keep the terms linear in $m_a$ in the numerators of the integrands in Eqs.~\eqref{eq:propdiagur} and Eq.~\eqref{eq:Dirpropvac}. However, in this situation, solving Eqs.~\eqref{eq:Dyseqab} and~\eqref{eq:Dyseqaa} would be more awkward especially in the matter case in Sec.~\ref{subsec:DRESSPROPMATT}.

That is why we have chosen an alternative way for the regularization by introducing the parameters $\xi$ and $\alpha_a$ in Sec.~\ref{subsec:DRESSPROPMATT}, as well as $\varsigma_a$ in Sec.~\ref{subsec:DRESSPROPB}. Note that $\alpha_a = (p+g_{a}/2)/E_{a-}$ and $\varsigma_a = p/E_a$ have the physical meaning. The quantity $\xi$ was introduced formally. All these parameters are supposed to approach to one for ultrarelativistic neutrinos.

We made at least one consistency check for the chosen regularization in Secs.~\ref{subsec:DRESSPROPMATT} and~\ref{subsec:DRESSPROPB}. We have demonstrated that the diagonal dressed propagators coincide with the undressed ones if one turns off the nondiagonal neutrino interactions, i.e. at $g\to 0$ and $\mu\to 0$ (see the discussion below Eqs.~\eqref{eq:Sigmaaa} and~\eqref{eq:Sigma22Bfin}).

\section*{Acknowledgments}

I am thankful to M.~Deka for the help with the Feynman diagram drawing.

\appendix

\section{Propagators of Dirac neutrinos in matter}\label{sec:DIRPROPMATT}

In this Appendix, we derive the diagonal propagator of a massive Dirac
neutrino interacting with background matter. The nondiagonal matter
interaction, described by the effective potential $G$, is not accounted
for here. Thus, the propagators considered in this Appendix correspond
to thin lines in Fig.~\ref{fig:feyndiag}.

The mass eigenstate $\psi_{a}$ of a Dirac neutrino, interacting with
a nonmoving and unpolarized background matter, obeys the wave equation,
\begin{equation}\label{eq:Direqmattdiag}
  \mathrm{i}\dot{\psi}_{a}=
  \left(
    \bm{\alpha}\mathbf{p}+\beta m_{a}+g_{a}\frac{1-\gamma^{5}}{2}
  \right)
  \psi_{a},
\end{equation}
where $m_{a}$ is the mass of $\psi_{a}$, $g_{a}$ is the diagonal
effective potential of the matter interaction, $\bm{\alpha}=\gamma^{0}\bm{\gamma}$
and $\beta=\gamma^{0}$ are the Dirac matrices, and $\mathbf{p}=-\mathrm{i}\nabla$
is the momentum operator. Equation~(\ref{eq:Direqmattdiag}) can
be obtained from Eqs.~\eqref{eq:massLagr} and~(\ref{eq:Direqmattmass})
if we omit there the nondiagonal potential, $g=0$.

One can check that the helicity operator $\bm{\Sigma}\mathbf{p}$
commutes with the Hamiltonian of Eq.~(\ref{eq:Direqmattdiag}) provided
that $g_{a}$ is spatially uniform. Thus, the general solution of
Eq.~(\ref{eq:Direqmattdiag}) has the form,
\begin{equation}\label{eq:psiamatt}
  \psi_{a}(\mathbf{x},t)=\int\frac{\mathrm{d}^{3}p}{(2\pi)^{3/2}}e^{-\mathrm{i}g_{a}t/2}\sum_{\sigma=\pm}
  \left(
    a_{a\sigma}(\mathbf{p})u_{a\sigma}(\mathbf{p})e^{-\mathrm{i}E_{a\sigma}t+\mathrm{i}\mathbf{px}}
    +b_{a\sigma}^{\dagger}(\mathbf{p})v_{a\sigma}(\mathbf{p})e^{\mathrm{i}E_{a\sigma}t-\mathrm{i}\mathbf{px}}
  \right),
\end{equation}
where
\begin{equation}\label{eq:enlevmatt}
  E_{a\sigma}=\sqrt{\left(p-\sigma\frac{g_{a}}{2}\right)^{2}+m_{a}^{2}},
\end{equation}
are the energy levels~\cite{energynuantinu},
\begin{align}\label{eq:basismatt}
  u_{a\sigma}(\mathbf{p}) = & \sqrt{\frac{E_{a\sigma}-\sigma p+g_{a}/2}{2E_{\sigma}}}
  \left(
    \begin{array}{c}
      -\frac{m_{a}}{E_{a\sigma}-\sigma p+g_{a}/2}w_{\sigma}(\mathbf{p}) \\
      w_{\sigma}(\mathbf{p})
    \end{array}
  \right),
  \notag
  \\
  v_{a\sigma}(\mathbf{p}) = & \sqrt{\frac{E_{a\sigma}-\sigma p+g_{a}/2}{2E_{a\sigma}}}
  \left(
    \begin{array}{c}
      w_{-\sigma}(\mathbf{p})\\
      \frac{m_{a}}{E_{a\sigma}-\sigma p+g_{a}/2}w_{-\sigma}(\mathbf{p})
    \end{array}
  \right),
\end{align}
are the basis spinors which are normalized to one, $|u_{a\sigma}(\mathbf{p})|^{2}=|v_{a\sigma}(\mathbf{p})|^{2}=1$,
$b_{a\sigma}^{\dagger}(\mathbf{p})$ and $a_{a\sigma}(\mathbf{p})$
are the creation and annihilation operators for antineutrinos and
neutrinos, and
\begin{equation}\label{eq:helamp}
  w_{+}(\mathbf{p})=
  \left(
    \begin{array}{c}
      e^{-\mathrm{i}\varphi/2}\cos\vartheta/2\\
      e^{\mathrm{i}\varphi/2}\sin\vartheta/2
    \end{array}
  \right),
  \quad
  w_{-}(\mathbf{p})=
  \left(
    \begin{array}{c}
      -e^{-\mathrm{i}\varphi/2}\sin\vartheta/2\\
      e^{\mathrm{i}\varphi/2}\cos\vartheta/2
    \end{array}
  \right),
\end{equation}
are the helicity amplitudes. Here, $\vartheta$ and $\varphi$ are
spherical angles fixing the direction of $\mathbf{p}$. Equation~(\ref{eq:basismatt})
implies that the Dirac matrices are in the chiral representation.

Assuming that
\begin{equation}
  \left\{
    a_{a\sigma}(\mathbf{p}),a_{a\sigma'}^{\dagger}(\mathbf{q})
  \right\}_+ =
  \left\{
    b_{a\sigma}(\mathbf{p}),b_{a\sigma'}^{\dagger}(\mathbf{q})
  \right\}_+ =
  \delta_{\sigma\sigma'}\delta(\mathbf{p}-\mathbf{q}),
\end{equation}
with the rest of anticommutators being equal to zero, one gets that
$\left\{ \psi(\mathbf{x},t),\psi^{\dagger}(\mathbf{y},t)\right\}_+ =\delta(\mathbf{x}-\mathbf{y})$.
Thus, $\psi_{a}$ in Eq.~(\ref{eq:psiamatt}) is the properly quantized
solution of Eq.~(\ref{eq:Direqmattdiag}).

Defining the propagator of $\psi_{a}$ in the standard manner
\begin{equation}\label{eq:propmattdef}
  \mathrm{i}S_{a}(x-y)=\theta(x_{0}-y_{0})
  \left\langle
    0
    \left|
      \psi_{a}(x)\bar{\psi}_{a}(y)
    \right|
    0
  \right\rangle -
  \theta(y_{0}-x_{0})
  \left\langle
    0
    \left|
      \bar{\psi}_{a}(y)\psi_{a}(x)
    \right|
    0
  \right\rangle,
\end{equation}
and using Eqs.~(\ref{eq:psiamatt})-(\ref{eq:helamp}), we cast Eq.~(\ref{eq:propmattdef})
to the form,
\begin{align}\label{eq:propdiagmatt}
  S_{a}(x)= & \int\frac{\mathrm{d}^{4}p}{(2\pi)^{4}}e^{-\mathrm{i}px}\sum_{\sigma}\frac{1}{4E_{a\sigma}}
  \notag
  \\
  & \times
  \bigg[
    \left(
      m_{a}-\mathrm{i}(\sigma p-g_{a}/2)
      \sigma_2
    \right)
    \left(
      \frac{1}{p_{0}-E_{a\sigma}-g_{a}/2+\mathrm{i}0}-\frac{1}{p_{0}+E_{a\sigma}-g_{a}/2-\mathrm{i}0}
    \right)
    \nonumber
    \\
    & -
    E_{a\sigma}
    \sigma_1
    \left(
      \frac{1}{p_{0}-E_{a\sigma}-g_{a}/2+\mathrm{i}0}+\frac{1}{p_{0}+E_{a\sigma}-g_{a}/2-\mathrm{i}0}
    \right)
  \bigg]
  \otimes
  \left[
    1+\sigma
    \left(
      \bm{\sigma}\hat{\mathbf{p}}
    \right)
  \right],
\end{align}
where $\bm{\sigma}$ are the Pauli matrices and $\mathrm{i}0$ is
a small imaginary quantity. In Eq.~(\ref{eq:propdiagmatt}), we use
the agreement that, e.g.,
\begin{equation}
  \sigma_1 \otimes \bm{\sigma}
  \equiv
  \left(
    \begin{array}{cc}
      0 & \bm{\sigma}\\
      \bm{\sigma} & 0
    \end{array}
  \right),
\end{equation}
etc.

Note that, if we turn off the neutrino matter interaction, i.e. we
put $g_{a}=0$, we rewrite Eq.~(\ref{eq:propdiagmatt}) in the form,
\begin{equation}\label{eq:vacpropDirgen}
  S_{a}(x)=\int\frac{\mathrm{d}^{4}p}{(2\pi)^{4}}e^{-\mathrm{i}px}\frac{\gamma^{\mu}p_{\mu}+m_{a}}{p^{2}-m_{a}^{2}+\mathrm{i}0},
\end{equation}
which coincides with the usual propagator of a massive Dirac field
in vacuum.

While studying flavor oscillations in matter in Sec.~\ref{subsec:DRESSPROPMATT},
we neglect antineutrino contributions and consider ultrarelativistic
neutrinos, which are left polarized. It means that, in Eq.~(\ref{eq:propdiagmatt}),
we should neglect the terms containing $-\mathrm{i}0$ in the denominators,
the neutrino mass in the numerator, and the terms corresponding to
$\sigma=+1$. The matrix $1-\left(\bm{\sigma}\hat{\mathbf{p}}\right)$ turns
out to be singular. That is why we regularize it by $1-\xi\left(\bm{\sigma}\hat{\mathbf{p}}\right)$,
where $\xi\to1$. Finally, we get the propagator of a massive ultrarelativistic
neutrino in matter in the momentum space,
\begin{equation}\label{eq:propdiagur}
  S_{a}(p_{0},\mathbf{p})=
  -\frac{1}{4(p_{0}-E_{a-}-g_{a}/2+\mathrm{i}0)}(1-\alpha_{a}\gamma^{5})
  \left(
    \begin{array}{cc}
      0 & 1-\xi
      \left(
        \bm{\sigma}\hat{\mathbf{p}}
      \right)\\
      1-\xi
      \left(
        \bm{\sigma}\hat{\mathbf{p}}
      \right) & 0
    \end{array}
  \right),
\end{equation}
where $\alpha_{a}=(p+g_{a}/2)/E_{a-}$. Using Eq.~(\ref{eq:enlevmatt}),
one gets that $\alpha_{a}\to1$ for ultrarelativistic neutrinos. However,
we keep $\alpha_{a}\neq1$, to avoid dealing with a singular projection
operator $1-\gamma^{5}$ in Eq.~(\ref{eq:propdiagur}).

\section{Computation of integrals}\label{sec:COMPINT}

In this Appendix, we provide some details for the computation of integrals
in the matrix elements in matter and a magnetic field.

The matrix element in Eq.~(\ref{eq:matrelcylcoord}) is expressed
in the form, $\mathcal{M}_{e\to\mu}=I_{+}-I_{-}$, where
\begin{align}\label{eq:Ipm}
  I_{\pm} = & \frac{1}{16\pi^{2}}\int_{0}^{\infty}\rho\mathrm{d}\rho\int_{-\infty}^{+\infty}\mathrm{d}z
  \left(
    1+\frac{z}{\sqrt{z^{2}+\rho^{2}}}
  \right)
  e^{\mathrm{i}zL}
  \notag
  \\
  & \times
  \frac{\frac{1}{2}(E_{2}-E_{1}+g_{2}-g_{1})\sin2\theta+g\cos2\theta}{(E-\mathcal{E}_{\pm}+\mathrm{i}0)\sqrt{
  \left(
    \frac{E_{2}-E_{1}+g_{2}-g_{1}}{2}
  \right)^{2}
  +g^{2}}},
\end{align}
We consider one of the integrals in Eq.~(\ref{eq:Ipm}), e.g., $I_{+}$.

The integration over $z$ is characterized by the poles, which are
the roots of the equation $E-\mathcal{E}_{+}+\mathrm{i}0=0$. One
can show that these roots are in the upper half-plane. Nevertheless,
it is not an easy task to find all the roots analytically. We find
an approximate solution of this equation for ultrarelativistic neutrinos. The roots
are $z_{+}=\sqrt{\rho_{0}^{2}-\rho^{2}} + \mathrm{i}0$ if $\rho<\rho_{0}$, and
$z_{+}=\mathrm{i}\sqrt{\rho^{2}-\rho_{0}^{2}}$ if $\rho>\rho_{0}$,
where $\rho_{0}=\bar{E}_{m}+\mathfrak{E}_{m}$, with $\bar{E}_{m}$
and $\mathfrak{E}_{m}$ given in Eq.~(\ref{eq:Em}).

After the integration over $z$, we rewrite $I_{+}$ for ultrarelativistic
neutrinos in the form,
\begin{align}\label{eq:I+}
  I_{+}= & -\frac{\mathrm{i}\bar{E}_{m}}{8\pi}\frac{
  \left(
    \frac{\Delta m^{2}}{4E}+\frac{g_{2}-g_{1}}{2}
  \right)
  \sin2\theta+g\cos2\theta}{\sqrt{
  \left(
    \frac{\Delta m^{2}}{4E}+\frac{g_{2}-g_{1}}{2}
  \right)^{2}+g^{2}}}I_{\rho},
  \quad
  I_{\rho}=\int_{0}^{\infty}\rho\mathrm{d}\rho
  \left(
    1+\frac{z_{+}}{\rho_{0}}
  \right)
  \frac{e^{\mathrm{i}z_{+}L}}{z_{+}},
\end{align}
The remaining integral over $\rho$ in Eq.~(\ref{eq:I+}) is computed
explicitly,
\begin{align}\label{eq:Irho}
  I_{\rho}= & \int_{0}^{\rho_{0}}\rho\mathrm{d}\rho
  \left(
    1+\frac{\sqrt{\rho_{0}^{2}-\rho^{2}}}{\rho_{0}}
  \right)
  \frac{e^{\mathrm{i}\sqrt{\rho_{0}^{2}-\rho^{2}}L}}{\sqrt{\rho_{0}^{2}-\rho^{2}}}-\mathrm{i}\int_{\rho_{0}}^{\infty}\rho\mathrm{d}\rho
  \left(
    1+\frac{\mathrm{i}\sqrt{\rho^{2}-\rho_{0}^{2}}}{\rho_{0}}
  \right)
  \frac{e^{-\sqrt{\rho^{2}-\rho_{0}^{2}}L}}{\sqrt{\rho^{2}-\rho_{0}^{2}}}
  \nonumber
  \\
  & =
  -\frac{2\mathrm{i}e^{\mathrm{i}\rho_{0}L}}{L}
  \left(
    1+\frac{\mathrm{i}}{2\rho_{0}L}
  \right).
\end{align}
%
To obtain Eq.~\eqref{eq:Irho} we use the values of the following definite integrals:
\begin{align}\label{eq:basint}
  \int_{0}^{\rho_{0}}\rho\mathrm{d}\rho\frac{e^{\mathrm{i}L\sqrt{\rho_{0}^{2}-\rho^{2}}}}{\sqrt{\rho_{0}^{2}-\rho^{2}}} & =
  -\frac{\mathrm{i}}{L}(e^{\mathrm{i}\rho_{0}L}-1),
  \notag
  \\
  \int_{0}^{\rho_{0}}\rho\mathrm{d}\rho e^{\mathrm{i}L\sqrt{\rho_{0}^{2}-\rho^{2}}} & =
  -\frac{\mathrm{i}}{L^{2}}e^{\mathrm{i}\rho_{0}L}(L\rho_{0}+\mathrm{i})-\frac{1}{L^{2}},
  \notag
  \\
  \int_{0}^{\rho_{0}}\rho^3 \mathrm{d}\rho
  \frac{e^{\mathrm{i}L\sqrt{\rho_{0}^{2}-\rho^{2}}}}{\sqrt{\rho_{0}^{2}-\rho^{2}}} & =
  -\frac{2}{L^{3}}e^{\mathrm{i}\rho_{0}L}(L\rho_{0}+\mathrm{i})
  +\frac{\mathrm{i}}{L^{3}}(L^2 \rho_0^2 + 2),
  \notag
  \\
  \int_{\rho_{0}}^{\infty}\rho\mathrm{d}\rho\frac{e^{-L\sqrt{\rho^{2}-\rho_{0}^{2}}}}{\sqrt{\rho^{2}-\rho_{0}^{2}}} & =
  \frac{1}{L},
  \notag
  \\
  \int_{\rho_{0}}^{\infty}\rho\mathrm{d}\rho e^{-L\sqrt{\rho^{2}-\rho_{0}^{2}}} & =
  \frac{1}{L^{2}},
  \notag
  \\
  \int_{\rho_{0}}^{\infty}\rho^3\mathrm{d}\rho \frac{e^{-L\sqrt{\rho^{2}-\rho_{0}^{2}}}}{\sqrt{\rho^{2}-\rho_{0}^{2}}} & =
  \frac{2}{L^{3}} + \frac{\rho_0^2}{L}. 
\end{align}
As a rule, one assumes that the propagation length is long enough, $L\gg\bar{E}_{m}^{-1}$. However, we keep the terms $\propto L^{-2}$ to study the behavior of the probabilities at relatively short distances.

Equation~(\ref{eq:Irho}) allows one to finalize the computation
of $I_{+}$ in Eq.~(\ref{eq:I+}). The expression for $I_{-}$ can
be found analogously. We provide just the final results for both integrals,
\begin{equation}\label{eq:Ipnfin}
  I_{\pm}=-\frac{\bar{E}_{m}e^{\mathrm{i}(\bar{E}_{m}\pm\mathfrak{E}_{m})L}}{4\pi L}\frac{
  \left(
    \frac{\Delta m^{2}}{4E}+\frac{g_{2}-g_{1}}{2}
  \right)
  \sin2\theta+g\cos2\theta}{\sqrt{
  \left(
    \frac{\Delta m^{2}}{4E}+\frac{g_{2}-g_{1}}{2}
  \right)^{2}
  +g^{2}}}
  \left(
    1 + \frac{\mathrm{i}}{2(\bar{E}_m\pm \mathfrak{E}_{m})L}
  \right).
\end{equation}
Equation~(\ref{eq:Ipnfin}) is used to derive Eq.~(\ref{eq:Memug}).

Now, we compute the matrix element for the spin-flavor precession
in a magnetic field in Eq.~(\ref{eq:matrelSFPgen}). We also decompose
it as $\mathcal{M}_{e\to\mu}=I_{+}-I_{-}$. Using the same cylindrical
coordinates as in the matter case above (see also Sec.~\ref{subsec:QFTMSW}),
one rewrites $I_{\pm}$ in the form,
\begin{align}\label{eq:IpmB}
  I_{\pm} = & \frac{\sin2\theta}{64\pi^{3}}
  \int_{0}^{2\pi}\mathrm{d}\phi\int_{0}^{\infty}\rho\mathrm{d}\rho\int_{-\infty}^{+\infty}\mathrm{d}z
  \left(
    1+\frac{z}{\sqrt{z^{2}+\rho^{2}}}
  \right)
  e^{\mathrm{i}zL}
  \notag
  \\
  & \times
  \frac{(E_{2}-E_{1})}{(E-\mathcal{E}_{\pm}+\mathrm{i}0)\sqrt{
  \left(
    \frac{E_{2}-E_{1}}{2}
  \right)^{2}
  +\mu^{2}\tilde{B}^{2}}},
\end{align}
where 
\begin{equation}
  \mathcal{E}_{\pm}=\bar{E}\pm\sqrt{
  \left(
    \frac{E_{2}-E_{1}}{2}
  \right)^{2}
  +\mu^{2}\tilde{B}^{2}},
  \quad
  \bar{E}=\frac{E_{1}+E_{2}}{2},
\end{equation}
and $\tilde{B}^{2}=B^{2}\left(1-\frac{\rho^{2}\cos^{2}\phi}{\rho^{2}+z^{2}}\right)$.

Considering again $I_{+}$ only, one gets from Eq.~(\ref{eq:IpmB})
that
\begin{equation}\label{eq:I+B}
  I_{+}=-\frac{\mathrm{i}\Delta m^{2}\sin2\theta}{64\pi^{2}\mathfrak{E}_{\mathrm{B}}}
  \int_{0}^{2\pi}\mathrm{d}\phi\int_{0}^{\infty}\rho\mathrm{d}\rho\frac{e^{\mathrm{i}z_{+}L}}{z_{+}}
  \left(
    1+\frac{z_{+}}{E+\mathfrak{E}_{\mathrm{B}}}\right)\left(1+\frac{(\mu B)^{2}\rho^{2}\cos^{2}\phi}{2E^{2}\mathfrak{E}_{\mathrm{B}}^{2}}
  \right).
\end{equation}
Here
\begin{equation}
  z_{+}=\sqrt{1+\frac{(\mu B)^{2}}{E\mathfrak{E}_{\mathrm{B}}}\cos^{2}\phi}\times
  \begin{cases}
    \sqrt{\rho_{0}^{2}-\rho^{2}} + \mathrm{i}0, & \text{if}\ \rho<\rho_{0},\\
    \mathrm{i}\sqrt{\rho^{2}-\rho_{0}^{2}}, & \text{if}\ \rho>\rho_{0},
  \end{cases}
\end{equation}
where
\begin{equation}
  \rho_{0}=\frac{E+\mathfrak{E}_{\mathrm{B}}}{\sqrt{1+\frac{(\mu B)^{2}}{E\mathfrak{E}_{\mathrm{B}}}\cos^{2}\phi}},
\end{equation}
and $\mathfrak{E}_{\mathrm{B}}$ is given in Eq.~(\ref{eq:EB}).

The calculation of the integral over $\rho$ in Eq.~(\ref{eq:I+B}) is analogous
to Eq.~(\ref{eq:Irho}). We split the integration segment as $[0,\infty)=[0,\rho_{0}]\cup[\rho_{0},\infty)$ and use Eq.~\eqref{eq:basint}. Eventually,
one gets that
\begin{align}\label{eq:I+Bphi}
  I_{+}= & -\frac{\mathrm{i}\Delta m^{2}\sin2\theta}{64\pi^{2}\mathfrak{E}_{\mathrm{B}}L}e^{\mathrm{i}(E+\mathfrak{E})L}
  \int_{0}^{2\pi}\frac{\mathrm{d}\phi}{1+\frac{(\mu B)^{2}}{E\mathfrak{E}_{\mathrm{B}}}\cos^{2}\phi}
  \nonumber
  \\
  & \times
  \left\{
    2+\frac{\mathrm{i}}{(E+\mathfrak{E}_{\mathrm{B}})L}-\frac{\mathrm{i}(\mu B)^{2}\cos^{2}\phi}{E^{2}\mathfrak{E}_{\mathrm{B}}^{2}L^{2}
    \left(
      1+\frac{(\mu B)^{2}}{E\mathfrak{\mathfrak{E}_{\mathrm{B}}}}\cos^{2}\phi
    \right)}
    \left[
      (E+\mathfrak{E}_{\mathrm{B}})L+\mathrm{i}
    \right]
  \right\}.
\end{align}
Then, we calculate the remaining integral
over $\phi$ in Eq.~(\ref{eq:I+Bphi}) in the approximation $(\mu B)^{2}\ll E\mathfrak{E}_{\mathrm{B}}$.
We present the final result for both $I_{+}$ and $I_{-}$, which
is calculated in a similar manner,
\begin{equation}\label{eq:IpmBfin}
  I_{\pm}=-\frac{Ee^{\mathrm{i}(E\pm\mathfrak{E}_{\mathrm{B}})L}}{4\pi L}\frac{\sin2\theta\frac{\Delta m^{2}}{4E}}{\sqrt{(\mu B)^{2}+
  \left(
    \frac{\Delta m^{2}}{4E}
  \right)^{2}}}
  \left(
    1-\frac{(\mu B)^{2}}{2E\mathfrak{E}_{\mathrm{B}}}
  \right)
  \left[  
    1 +\frac{\mathrm{i}}{2EL}
    \left(
      1-\frac{(\mu B)^{2}}{2\mathfrak{E}_\mathrm{B}^{2}}
    \right)
  \right].
\end{equation}
Analogously to Eq.~\eqref{eq:Ipnfin}, we keep only the lext-to-leading term $\propto L^{-2}$ in Eq.~\eqref{eq:IpmBfin}. We use Eq.~(\ref{eq:IpmBfin}) in Eq.~(\ref{eq:martelB}).

\section{Propagators of Dirac neutrinos in vacuum}\label{sec:DIRPROPVAC}

While studying the spin-flavor precession of neutrinos with a transition
magnetic moment, the undressed propagators do not involve the neutrino
magnetic interaction. Thus, this kind of propagators is the vacuum
one. In this Appendix, we derive the propagators and express them
in the appropriate form used in Sec.~\ref{subsec:DRESSPROPB}.

The propagators in question can be obtained directly from Eq.~(\ref{eq:vacpropDirgen}).
For this purpose, we rewrite Eq.~(\ref{eq:vacpropDirgen}) in the
equivalent form,
\begin{align}\label{eq:Dirproptransf}
  S_{a}(x) = & \frac{1}{2}\int\frac{\mathrm{d}^{4}p}{(2\pi)^{4}}e^{-\mathrm{i}px}
  \bigg[
    \frac{1}{p_{0}-E_{a}+\mathrm{i}0}
    \left(
      \gamma^{0}-\frac{p}{E_{a}}\bm{\gamma}\hat{\mathbf{p}}+\frac{m_{a}}{E_{a}}
    \right)
    \notag
    \\
    & +
    \frac{1}{p_{0}+E_{a}-\mathrm{i}0}
    \left(
      \gamma^{0}+\frac{p}{E_{a}}\bm{\gamma}\hat{\mathbf{p}}-\frac{m_{a}}{E_{a}}
    \right)
  \bigg],
\end{align}
where $E_{a}=\sqrt{p^{2}+m_{a}^{2}}$ is the energy of a neutrino
mass eigenstate in vacuum.

If we deal with ultrarelativistic neutrinos, in Eq.~(\ref{eq:Dirproptransf}),
we should neglect the term containing $-\mathrm{i}0$ in the denominator,
as well as the small term $m_{a}/E_{a}$ in the numerator, analogously
to Appendix~\ref{sec:DIRPROPMATT}. Thus, one gets from Eq.~(\ref{eq:Dirproptransf})
that
\begin{equation}\label{eq:Dirpropvac}
  S_{a}(p_{0},\mathbf{p})\to\frac{\gamma^{0}-\varsigma_{a}\bm{\gamma}\hat{\mathbf{p}}}{2(p_{0}-E_{a}+\mathrm{i}0)},
\end{equation}
where $\varsigma_{a}=p/E_{a}$. For ultrarelativistic particles, $\varsigma_{a}\to1$.
However, we keep $\varsigma_{a}\neq1$ in Eq.~(\ref{eq:Dirpropvac})
to avoid the singularity in $S_{a}$. We set $\varsigma_{a}\to1$
in the final expression for the dressed propagators in Sec.~\ref{subsec:DRESSPROPB}.


\begin{thebibliography}{50}

\bibitem{Abe25}
  K.~Abe et al.
  (The NOvA Collaboration \& The T2K Collaboration),
  Joint neutrino oscillation analysis from the T2K and NOvA experiments,
  Nature \textbf{646}, 818--824 (2025)
  [arXiv:2510.19888].

\bibitem{Wol78}
  L.~Wolfenstein,
  Neutrino Oscillations in Matter,
  Phys. Rev. D \textbf{17}, 2369--2374 (1978).

\bibitem{MikSmi85}
  S.~P.~Mikheyev and A.~Yu.~Smirnov,
  Resonance Amplification of Oscillations in Matter and Spectroscopy of Solar Neutrinos,
  Sov. J. Nucl. Phys. \textbf{42}, 913--917 (1985).

\bibitem{CheXu25}
  S.~Chen and X.-J.~Xu,
  Solar neutrinos,
  arXiv:2501.09971.

\bibitem{LeeSch77}
  B.~W.~Lee and R.~Shrock,
  Natural suppression of symmetry violation in gauge theories: Muon- and electron-lepton-number nonconservation,
  Phys. Rev. D \textbf{16}, 1444--1473 (1977).

\bibitem{DvoStu04}
  M.~Dvornikov and A.~Studenikin,
  Electric charge and magnetic moment of a massive neutrino,
  Phys. Rev. D \textbf{69}, 073001(2004)
  [hep-ph/0305206].

\bibitem{LimMar88}
  C.-S.~Lim and W.~J.~Marciano,
  Resonant spin-flavor precession of solar and supernova neutrinos,
  Phys. Rev. D \textbf{37}, 1368--1373 (1988).

\bibitem{Akh88}
  E.~Kh.~Akhmedov,
  Resonant Amplification of Neutrino Spin Rotation in Matter and the Solar Neutrino Problem,
  Phys. Lett. B \textbf{213}, 64--68 (1988).

\bibitem{Kin04}
  S.~F.~King,
  Neutrino Mass Models,
  Rept. Prog. Phys. \textbf{67}, 107--158 (2004)
  [hep-ph/0310204].

\bibitem{Agr25}
  A.~Agrawal et al. (The AMoRE Collaboration),
  Improved Limit on Neutrinoless Double Beta Decay of $^{100}\text{Mo}$ from AMoRE-I,
  Phys. Rev. Lett. \textbf{134}, 082501 (2025)
  [arXiv:2407.05618].

\bibitem{Giu25}
  C.~Giunti, K.~Kouzakov, Y.-F.~Li, and A.~Studenikin,
  Neutrino Electromagnetic Properties,
  Ann. Rev. Nucl. Part. Sci. \textbf{75}, 1--33 (2025)
  [arXiv:2411.03122].

\bibitem{NauNau20}
  D.~V.~Naumov and V.~A.~Naumov,
  Quantum Field Theory of Neutrino Oscillations,
  Phys. Part. Nucl. \textbf{51}, 1--106 (2020).

\bibitem{Kob82}
  I.~Yu.~Kobzarev, B.~V.~Martem'yanov, L.~B.~Okun', and M.~G.~Shchepkin,
  Sum rules for neutrino oscillations,
  Sov. J. Nucl. Phys. \textbf{35}, 708--712 (1982).

\bibitem{GiuKimLee93}
  C.~Giunti, C.~W.~Kim, and J.~A.~Lee,
  On the treatment of neutrino oscillations without resort to weak eigenstates
  Phys. Rev. D \textbf{48}, 4310--4317 (1993)
  [hep-ph/9305276].

\bibitem{GriSto96}
  W.~Grimus and P.~Stockinger,
  Real oscillations of virtual neutrinos,
  Phys. Rev. D \textbf{54}, 3414--3419 (1996)
  [hep-ph/9603430].

\bibitem{EgoVol22}
  V.~Egorov and I.~Volobuev,
  Quantum field-theoretical description of neutrino oscillations in magnetic field,
  J. Exp. Theor. Phys. \textbf{135}, 197--208 (2022)
  [arxiv:2107.11570].

\bibitem{Dvo25}
  M.~Dvornikov,
  Quantum field theory treatment of neutrino flavor oscillations in matter,
  Phys. Rev. D \textbf{111}, 056009 (2025)
  [arxiv:2411.19120].

\bibitem{Dvo25b}
  M.~Dvornikov,
  Quantum field theory treatment of the neutrino spin-flavor precession in a magnetic field,
  arxiv:2504.14726.

\bibitem{CarChu99}
  C.~Y.~Cardall and D.~J.~H.~Chung,
  MSW effect in quantum field theory,
  Phys. Rev. D \textbf{60}, 073012 (1999)
  [hep-ph/9904291].

\bibitem{Dvo11}
  M.~Dvornikov,
  Field theory description of neutrino oscillations,
  in \textit{Neutrinos: Properties, Sources and Detection},
  ed. by J.~P.~Greene (Nova Science Publishers, New York, 2011), pp.~23--90
  [arxiv:1011.4300].

\bibitem{AkhWil13}
  E.~Kh.~Akhmedov and A.~Wilhelm,
  Quantum field theoretic approach to neutrino oscillations in matter,
  J. High Energy Phys. \textbf{01}, 165 (2013)
  [arXiv:1205.6231].

\bibitem{NauShk21}
  V.~A.~Naumov and D.~S.~Shkirmanov,
  Reactor Antineutrino Anomaly Reanalysis in Context of Inverse-Square Law Violation,
  Universe \textbf{7}, 246 (2021).

\bibitem{ZhaQiaFal24}
  C.~Zhang, X.~Qian, and M.~Fallot,
  Reactor antineutrino flux and anomaly,
  Prog. Part. Nucl. Phys. \textbf{136}, 104106 (2024)
  [arXiv:2310.13070].

\bibitem{millichnu}
  We do not consider an exotic situation when neutrinos can be millicharged particles (see, e.g., Ref.~\cite{Giu25}).

\bibitem{Pas00}
  S.~Pastor, J.~Segura, V.~B.~Semikoz, and J.~W.~F.~Valle,
  A potential test of the CP properties and Majorana nature of neutrinos,
  Nucl. Phys. B \textbf{566}, 92--102 (2000)
  [hep-ph/9905405].

\bibitem{GlaIliMai70}
  S.~L.~Glashow, J.~Iliopoulos, and L.~Maiani,
  Weak Interactions with Lepton-Hadron Symmetry,
  Phys. Rev. D \textbf{2}, 1285--1292 (1970).

\bibitem{energynuantinu}
  Note that the total energy of neutrinos is $g_{a}/2+E_{a\sigma}$ and that of antineutrinos is $-g_{a}/2+E_{a,-\sigma}$.
 
\bibitem{KovSim24}
  S.~Kovalenko and F.~\v{S}imkovic, Neutrino oscillations as a single Feynman diagram,
  J. Phys. G: Nucl. Part. Phys. \textbf{51}, 035202 (2024)
  [Erratum: ibid. \textbf{51}, 059601 (2024)]. 
  
\bibitem{Bog90}
  N.~N.~Bogolubov, A.~A.~Logunov, A.~I.~Oksak, and I.~Todorov,
  \textit{General principles of quantum field theory}
  (Kluwer, Dordrecht, 1990), p.~270--317.

\bibitem{Wei95}
  S.~Weinberg,
  \textit{The Quantum Theory of Fields: Vol.~I. Foundations}
  (Cambridge University Press, Cambridge, 1995), p.~63.

\bibitem{Vol17}
  I.~P.~Volobuev,
  Quantum field-theoretical description of neutrino and neutral kaon oscillations,
  Int. J. Mod. Phys. A \textbf{33}, 1850075 (2018)
  [arXiv:1703.08070].

\bibitem{Beu03}
  M.~Beuthe,
  Oscillations of neutrinos and mesons in quantum field theory,
  Phys. Rept. \textbf{375}, 105--218 (2003)
  [hep-ph/0109119].

\bibitem{EgoVol19}
  V.~O.~Egorov and I.~P.~Volobuev,
  Coherence length of neutrino oscillations in quantum field-theoretical approach,
  Phys. Rev. D \textbf{100}, 033004 (2019)
  [arXiv:1902.03602].

\bibitem{DobMelSch25}
  I.~Dobrev, K.~Melnikov, and T.~Schwetz,
  Neutrino oscillations and scattering theory,
  J. High Energy Phys. \textbf{07}, 035 (2025)
  [arXiv:2504.10600].
  
\bibitem{GriStoMoh99}
  W.~Grimus, P.~Stockinger, and S.~Mohanty,
  The field theoretical approach to coherence in neutrino oscillations,
  Phys. Rev. D \textbf{59}, 013011 (1999)
  [hep-ph/9807442]. 

\bibitem{Dvo26}
  M.~Dvornikov,
  Impact of antiparticle degrees of freedom on neutrino flavor oscillations in frames of quantum field theory,
  to be published in Phys. Part. Nucl. (2026)
  [arxiv:2512.08361].

\end{thebibliography}
\end{document}